\documentclass[preprint,showpacs]{revtex4}
\usepackage{graphicx}
\usepackage{bm}
\usepackage{amsmath} 

\newcommand{\half}{\mbox{${\textstyle \frac{1}{2}}$}}           

\newcommand{\rd}{\textrm{d}}
\begin{document}

\title{Sequential hyperon decays in the reaction $e^+e^-\rightarrow\Sigma^0\bar{\Sigma}^0$}
\date{\today}
\author{G\"oran F\"aldt}\email{goran.faldt@physics.uu.se} 
\author{Karin Sch{\"o}nning}
\email{karin.schonning@physics.uu.se}
\affiliation{ Department of physics and astronomy, \
Uppsala University,
 Box 516, S-751 20 Uppsala,Sweden }


\begin{abstract}
We report on a study of the sequential hyperon decay 
$\Sigma^0\rightarrow\Lambda \gamma;\Lambda\rightarrow p\pi^- $ and its corresponding anti-hyperon 
decay. We derive a multi-dimentional and model-independent formalism for the case when 
 the hyperons are produced in the reaction $e^+e^-\rightarrow\Sigma^0\bar{\Sigma}^0$.
Cross-section distributions are calculated  using the folding technique. We also study 
sequential decays of single-tagged hyperons. 
\end{abstract}


\maketitle
%

%
%
 \date{\today} 
%

%
%
%
\section{Introduction}\label{ett}

The BESIII experiment \cite{Ablikim17a} has created new opportunities for research into  hyperon physics, based on $e^+e^-$ annihilation into hyperon-anti-hyperon pairs. 
Such possibilities 
are interesting, and for several reasons:
\begin{itemize}
    \item They offer the currently only feasible way for investigating the electromagnetic structure of hyperons \cite{pacetti}.
    \item By measuring in the vicinity of  vector-charmonium states, one gains information on     the strong baryon-antibaryon decay processes of charmonia.
    \item They offer a model-independent method for measuring weak-decay-asymmetry parameters, which can probe CP symmetry \cite{Nature}.
\end{itemize}

The basic reaction, $e^+e^-\rightarrow Y\bar{Y}$, is graphed in Fig.1. In the continuum region, \textit{i.e.}, in energy regions that do not overlap with energies of
 vector charmonia like $J/\psi$, $\psi'$ and $\psi(2S)$, the production process is dominated by one-photon exchange,  $e^+e^-\rightarrow \gamma^* \rightarrow Y\bar{Y}$. The reaction amplitude is then governed by the electromagnetic form factors $G_E$ and $G_M$. In the vicinity of vector resonances, the electromagnetic form factors are replaced by hadronic form factors 
 $G^{\psi}_E$ and $G^{\psi}_M$. However, the shapes of the differential-cross-section distributions are the same in the two cases: all physics of the  production mechanism is contained within the form factors, or equivalently, the ratio of form-factor magnitudes,
$\alpha_\psi$, and the relative phase of form factors, $\Delta \Phi_\psi$.

Analyses of  joint-decay distributions of hyperons, such as 
$\Lambda(\rightarrow p\pi^-) \bar{\Lambda}(\rightarrow \bar{p}\pi^+)$, enables us
to determine some of the weak-interaction-decay parameters, $\alpha\beta\gamma$.

\begin{figure}[ht]
     \centerline{\scalebox{0.60}{ \includegraphics{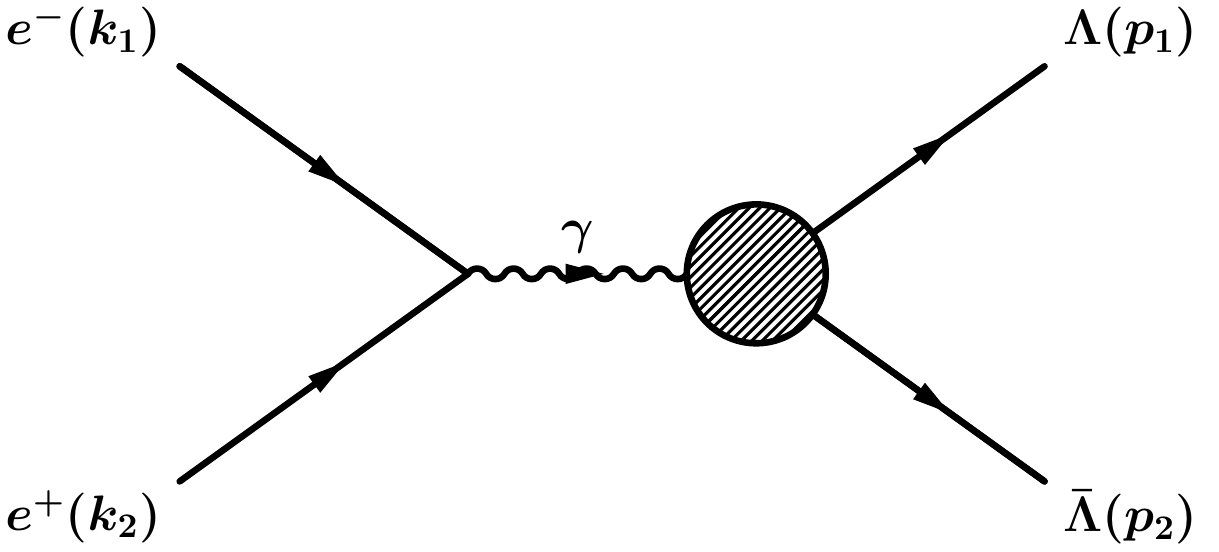} }}    
\caption{Graph describing the electromagnetic annihilation reaction  
$e^+ e^- \rightarrow  \bar{\Lambda} \Lambda$. The same reaction can also 
proceed hadronicly via  vector charmonium states such as 
$J/\psi$, $\psi'$, or $\psi(2{\textrm S})$,
 in place of the photon.}
\label{F1-fig}
\end{figure} 

The theoretical description of the annihilation reaction of Fig.1 is described 
in Ref.\cite{GF2}, and the corresponding annihilation reaction mediated by $J/\psi$ 
 in Ref.\cite{GF3}. Accurate experimental results for the form-factor 
parameters $\alpha_\psi$ and $\Delta\Phi_\psi$ and the 
weak-interaction parameters $\alpha_\Lambda(\alpha_{\bar{\Lambda}})$  for the latter 
annihilation process are all reported in Ref.\cite{Nature}.
A precise knowledge  of the asymmetry 
parameters $\alpha_\Lambda(\alpha_{\bar{\Lambda}})$ is
needed for studies  of  spin polarization in $\Omega^-$,
$\Xi^-$, and $\Lambda_c^+$ decays, and for tests of the standard model.

The graph of Fig.\ref{F1-fig} can be generalized in the sense that it can include hyperons that decay sequentially. It can also include cases where the produced hyperon is of a different type than the produced antihyperon, \textit{i.e.}, $e^+e^-\rightarrow Y_1\bar{Y_2}$.

In this note we shall consider annihilation into  $\Sigma^0\bar{\Sigma}^0$ pairs, Ref.\cite{GF4}. 
The $\Sigma^0$ decays electromagnetically, $\Sigma^0\rightarrow\Lambda\gamma$, and 
subsequently the 
Lambda hyperon  decays weakly, $\Lambda\rightarrow p\pi^-$. The interest of this measurement is many-fold:
\begin{itemize}
    \item The form factors provide information about the production process. So far, literature has focused on electromagnetic form factors whose interpretation is straight-forward 
		\cite{punjabi,pacetti}. However, recent experimental advances calls for an interpretation also of the hadronic form factors. In particular, it would be interesting to compare  the decay of $J/\psi$ into various hyperon-antihyperon pairs with the corresponding decays of other vector charmonia. 
    \item The BESIII collaboration plans to perform a first measurement of the branching fraction of the $\Sigma^0$ Dalitz decay $\Sigma^0 \to \Lambda \gamma^*, \gamma^* \to e^+e^-$ using the large data sample available for the $e^+ e^- \rightarrow  J/\psi \rightarrow \bar{\Sigma}^0 \Sigma^0$ process. Then, the most important background will come from $e^+ e^- \rightarrow  J/\psi \rightarrow \bar{\Sigma}^0 \Sigma^0;( \Sigma^0 \to \Lambda \gamma; \Lambda \to p\pi^- + c.c),$ where one of the photons undergoes external conversion into an $e^+e^-$ pair. This is because the branching ratio of the $\Sigma^0 \to \Lambda \gamma$, according to QED, is three orders of magnitude larger than that of the Dalitz decay. In order to properly account for the background, precise knowledge of the joint angular distribution is required.
    \item It can  provide an independent measurement of the Lambda asymmetry parameters
		 $\alpha_{\Lambda}$ and  $\alpha_{\bar{\Lambda}}$.
    \item It can provide a first test of strong CP symmetry in the $\Sigma^0 \to \Lambda \gamma$ decay Ref.\cite{St1}.
\end{itemize}		

Our calculation is performed in steps. 
First, we review some important facts; the spin structure of the
 $e^+e^-\rightarrow \Sigma^0\bar{\Sigma}^0$ annihilation reaction Ref.\cite{GF2}; 
the classical $\alpha\beta\gamma$ description of
hyperon decays Ref.\cite{Okun}; the 
description of the electromagnetic $\Sigma^0\rightarrow \Lambda \gamma$ decay, 
both for real and virtual photons Refs.\cite{Ber,GF4}. 
The virtual photons decay into Dalitz lepton pairs. An important element 
of our calculation is the factorization of the squared amplitudes 
 into a spin-independent fractional decay rate 
and a spin-density distribution.

Following these reviews  we demonstrate how the folding method of Ref.\cite{GF1} is adapted to  
sequential decays. Both simple and double decay chains are treated. Finally, we join 
production and decay steps to give the cross-section distributions.

The information we are hoping to gain resides in the angular distributions, and we are
therefore not overly concerned with absolute normalizations, although they may be obtained
without too much effort.

%
%
%
%
\section{Baryon form factors }\label{tvaa}

The diagram in Fig.1 describes the annihilation reaction
 $e^-(k_1)e^+(k_2)\rightarrow Y(p_1)\bar{Y}(p_2)$ and involves two 
vertex functions; one of them leptonic, the other one baryonic. The strength of the lepton-vertex function 
is determined by the electric charge $e_e$, but 
two form factors  $G_M(s)$ and $G_E(s)$ are needed for describing the baryonic vertex function. 
Here, $s=(p_1+p_2)^2$ with $p_1$ and $p_2$ as defined in Fig.1.

The strength of the baryon form factors is measured by the function $D(s)$, 
\begin{equation}
	D(s)=s\left| G_M \right|^2 + 4 M^2 \left| G_E \right|^2, \label{DS_def}
\end{equation}
a factor that multiplies all cross-section distributions. The ratio of  form factors
is measured by $\eta(s)$,
\begin{equation}
	\eta(s)= \frac{s\left| G_M \right|^2 - 4 M^2\left| G_E \right| ^2}
	{s\left| G_M\right|^2 + 4 M^2\left| G_E\right|^2},\label{alfa_def}
\end{equation}
with $\eta(s)$ satisfying $-1\leq \eta(s) \leq 1$.
The relative phase of form factors is measured by $\Delta\Phi(s)$, 
\begin{equation}
	\frac{G_E}{G_M}=e^{i\Delta\Phi(s)} \left| \frac{G_E}{G_M}\right|. \label{DPHI_def}
\end{equation}

In Ref.\cite{GF3} annihilation in the region of the $J/\psi$ and $\psi(2$S$)$ masses
 is considered.
The photon propagator of Fig.1 is then replaced by the appropriate vector-meson  propagator.



\section{Cross section for $e^-e^+\rightarrow Y(s_1) \bar{Y}(s_2)$} \label{trea}

Our first task is to review the calculation of the cross-section distribution for $e^+e^-$ annihilation 
into  baryon-antibaryon pairs, with baryon-four-vector polarizations
 $s_1$ and $s_2$ \cite{GF2,GF3}.
 From the squared matrix element of this process, $|{\cal{M}\,}|^2$, 
we remove  a factor $e_e^4/s^2$, which is the square of the propagator, and get 
\begin{equation}
	\rd \sigma= \frac{1}{2s} \,\frac{e_e^4}{s^2}\, \big| {\cal{M}}_{red}(s_1,s_2)\big|^2\,\,
	   \textrm{dLips}(k_1+k_2;p_1,p_2)	 ,\label{Basicsigma}
\end{equation}
with $s=(p_1+p_2)^2$, and $\textrm{dLips}$ the  phase-space element of Ref.\cite{Pil}.
 For a baryon 
of momentum $\mathbf{p}$ the four-vector 
spin $s$ is related to the three-vector spin $\mathbf{n}$, the spin in 
the rest system,  by
\begin{equation}
	s(\mathbf{p},\mathbf{n})=\frac{n_\|}{M}(|\mathbf{n}|,E\hat{\mathbf{p}})+(0,\mathbf{n}_\bot ).
	\label{4spin}
\end{equation}
Longitudinal and transverse directions of vectors are relative 
to the $\hat{\mathbf{p}}$ direction.

In the global 
c.m.~system kinematics simplifies. 
There, three-momenta $\mathbf{p}$ and $\mathbf{k}$ are defined such that
	\begin{align}
	\mathbf{p}_1  &=  - \mathbf{p}_2= \mathbf{p} , \\
	\mathbf{k}_1  &= - \mathbf{k}_2 = \mathbf{k}, 
\end{align}
	and with scattering angle $\theta$ defined by,
	\begin{equation}
		\cos\theta= \hat{\mathbf{p}}\cdot \hat{\mathbf{k}}.
	\end{equation}
	Furthermore, in the global 
c.m.~system the phase-space factor reads
\begin{equation}
	 \textrm{dLips}(k_1+k_2;p_1,p_2)	= \frac{p}{32\pi^2 k}\,\rd \Omega,
	\label{TwoDsig}
\end{equation}
with $p=|\mathbf{p}| $ and $k=|\mathbf{k}| $.

The matrix element in Eq.(\ref{Basicsigma}) can be written as a sum of  terms that depend 
on the baryon and antibaryon spin directions
in their respective rest systems,  $\mathbf{n}_1$ and $\mathbf{n}_2$, 
\begin{align}
	\left| {\cal{M}}_{red}(e^+e^-\right. &\rightarrow \left.Y(s_1)\overline{Y}(s_2))\right|^2= \, sD(s)\, 
	S(\mathbf{n}_{1},\mathbf{n}_{2}),\label{Hintro}
\end{align}
with the strength function $D(s)$ defined in Eq.(\ref{DS_def}). 
We call a function such as $S(\mathbf{n}_{1},\mathbf{n}_{2})$ a spin density. In the present case, the 
spin density is a sum of  seven mutually orthogonal contributions \cite{GF2},
\begin{align}
	S(\mathbf{n}_{1},\mathbf{n}_{2}) =& \,  {\cal{R}}  
	    + {\cal{S}}\, \mathbf{N}\cdot \mathbf{n}_{1}
	+ {\cal{S}}\,\mathbf{N}\cdot \mathbf{n}_{2}
	 + {\cal{T}}_1 
	  \mathbf{n}_{1} \cdot \hat{\mathbf{p}}\mathbf{n}_{2} \cdot \hat{\mathbf{p}}
	    \nonumber\\
	&  + {\cal{T}}_2  \mathbf{n}_{1\bot} \cdot \mathbf{n}_{{2}\bot} 
		+ {\cal{T}}_3 \mathbf{n}_{1\bot} \cdot \hat{\mathbf{k}}\mathbf{n}_{{2}\bot}
		\cdot \hat{\mathbf{k}}
		\nonumber\\
			&  + {\cal{T}}_4  \bigg( \mathbf{n}_{1} \cdot \hat{\mathbf{p}}\mathbf{n}_{{2}\bot} \cdot \hat{\mathbf{k}} 
			 + \mathbf{n}_{2} \cdot \hat{\mathbf{p}}\mathbf{n}_{1\bot} \cdot \hat{\mathbf{k}} 
			  \bigg)  ,\label{ISD}
\end{align}
where $\mathbf{N}$ is the normal to the scattering plane,
\begin{equation}
 \mathbf{N}=\frac{1}{\sin\theta} \,  \hat{\mathbf{p}}\times \hat{\mathbf{k}}.\label{Ndef}
\end{equation}

The seven structure functions ${\cal{R}}$, ${\cal{S}}$, and ${\cal{T}}$ of Eq.(\ref{ISD})  
 depend on the scattering angle $\theta$, the ratio function $\eta(s)$, and
the phase function $\Delta\Phi(s)$. Their detailed expressions are given in Appendix B.

The cross-section distribution for polarized final-state hyperons becomes
\begin{equation}
	\frac{\rd \sigma}{\rd \Omega}=\frac{p}{k} \frac{\alpha_e^2 \,D(s)}{4s^2}\,S(\mathbf{n}_{1},\mathbf{n}_{2}),
\end{equation}
where $\alpha_e$ is the fine-structure constant.

If we sum over baryon and antibaryon final-state polarizations we get a well-known result,
\begin{equation}
	\frac{\rd \sigma}{\rd \Omega}=\frac{p}{k} \frac{\alpha_e^2 \,D(s)}{s^2}\,
	{\cal{R}}  .\label{Compton-cc}
\end{equation}

Summing only over the antibaryon polarizations gives
\begin{equation}
	\frac{\rd \sigma}{\rd \Omega}=\frac{p}{k} \frac{\alpha_e^2 \,D(s)}{2s^2}\,({\cal{R}}  
	    + {\cal{S}}\, \mathbf{N}\cdot \mathbf{n}_{1}) .
\end{equation}
This result tells us that the baryon is polarized and that its polarization is
directed along the normal to the scattering plane, $ \hat{\mathbf{p}}\times 
	  \hat{\mathbf{k}}$, and that the value of the polarization is
\begin{equation}
	P_Y(\theta)=\frac{{\cal{S}}}{{\cal{R}}} =
\frac{\sqrt{1-{\eta}^2}\cos\theta\sin\theta}{1+\eta\cos^2\!\theta}\sin({\Delta\Phi})
\end{equation}
From Eq.(\ref{ISD}) we conclude that there is a corresponding result for the antibaryon,
 but it should then be remembered that $\mathbf{p}$ is the momentum of the  baryon 
but $-\mathbf{p}$  that of the antibaryon.

Baryon and antibaryon polarizations in $e^+e^-$ annihilation were first discussed by 
Dubni\u{c}kova {et al.}\cite{Dub}, but with results  slightly different from ours. 
For details see Ref.\cite{GF2}.
%
		
%
%
%
%
\section{Weak baryon decays}\label{fyra}

Weak decays of spin one-half baryons, such as $\Lambda\rightarrow p\pi^-$,
 involve two amplitudes, one S-wave
and one P-wave amplitude, and the decay distribution is commonly parametrized by  
three  parameters, denoted $\alpha\beta\gamma$,  and which fulfill a relation 
\begin{equation}
	\alpha^2 + \beta^2 +\gamma^2=1.
\end{equation}
Details of this description can be found in Refs.\cite{Lee} or \cite{Okun,GF2}.

Since we shall encounter several weak baryon decays of the same structure as the 
$\Lambda\rightarrow p\pi^-$ decay, we shall use a generic notation, $c\rightarrow d \pi$, 
for those decays.

The matrix element describing the decay of a polarized $c$ baryon into a polarized 
$d$ baryon  is 
\begin{equation}
	 {\cal{M}}(c\rightarrow d \pi)  =\bar{u}(p_d,s_d)(A+B\gamma_5)u(p_c,s_c), \label{HypMel}
\end{equation}
with $p$ and $s$ with appropriate indices denoting  momenta and spin four-vectors of
the baryons.
The square of this matrix element we factorize, writing
\begin{align}
	\big| {\cal{M}}(c\rightarrow d\pi)\big|^2 &=
	{\rm Tr}\Big[  \half(1+\gamma_5/\!\!\! s_d)(/\!\!\!p_d+m_d)(A+B\gamma_5) 
\nonumber\\
 & \times (/ \!\!\! p_c+m_c) \half(1+\gamma_5/\!\!\! s_c) (A^\star -B^\star \gamma_5)
\Big]. \nonumber\\ 
  &=R(c\rightarrow d\pi)\, G(\mathbf{n}_c,\mathbf{n}_d)
\label{MkvadWeak}
\end{align}
where $\mathbf{n}_c$ and $\mathbf{n}_d$ are the spin vectors of baryons $c$ and $d$ in
their rest frames, Eq.(\ref{4spin}). 
The $R$-factor is a spin independent factor,  defined by
\begin{align}
  R(c\rightarrow d\pi) &= 2m_c  \Gamma(c\rightarrow d\pi)/\Phi(c\rightarrow d\pi) ,
	 \nonumber\\
	&= \left|A\right|^2\left( (m_c+m_d)^2-m_\pi^2\right) \nonumber \\
	& \qquad \qquad+ \left|B\right|^2\left( (m_c-m_d)^2-m_\pi^2\right)
	,\label{Rrate}
\end{align}
where $\Phi(c\rightarrow d\pi)=\Phi(m_c;m_d,m_\pi)$ is the phase-space volume
of \ref{AppD}. We refer to $R(c\rightarrow d\pi)$ as the fractional decay rate, 
since it is a decay rate per unit phase space. Further
inspection of Eq.(\ref{MkvadWeak}) tells us, that 
$\Gamma(c\rightarrow d\pi)$ is defined as an average over the spins of
both initial and final baryons.

The  spin-density-distribution function,   
$G(\mathbf{n}_c,\mathbf{n}_d)$ of Eq.(\ref{MkvadWeak}), 
 is a Lorentz scalar, which we choose to evaluate 
in the rest system of the mother baryon, c,
\begin{equation}
	G(c, d) = 1+\alpha_c \mathbf{n}_c\cdot \mathbf{l}_d
	  +\alpha_c \mathbf{n}_d\cdot \mathbf{l}_d
		+\mathbf{n}_c\cdot \mathbf{L}_c(\mathbf{n}_d,\mathbf{l}_d ),
		\label{Gccd}
\end{equation}	
with
\begin{equation}
	\mathbf{L}_c(\mathbf{n}_d, \mathbf{l}_d)=\,\gamma_c \mathbf{n}_d
	+\bigg[(1-\gamma_c)\mathbf{n}_d\cdot \mathbf{l}_d\bigg]\, \mathbf{l}_d
	+\beta_c  \mathbf{n}_d\times \mathbf{l}_d .\label{Gccd_2}
\end{equation}
The vector $\mathbf{l}_d$ is a unit vector in the direction of motion
of the daughter baryon, $d$, in the rest system of mother baryon $c$. The indices on the 
$\alpha\beta\gamma$ 
parameters remind us they characterize  baryon $c$. A spin density is normalized 
if the spin-independent term is unity.

We observe an important  symmetry,
\begin{equation}
	\mathbf{n}_c\cdot \mathbf{L}_c(\mathbf{n}_d,\mathbf{l}_d )=
	\mathbf{n}_d\cdot \mathbf{L}_c(\mathbf{n}_c,-\mathbf{l}_d ).
\end{equation}

Since the spin of baryon $d$ is usually not measured, the interesting spin-density  is 
obtained by taking the
 average over the spin directions $\mathbf{n}_d$,
\begin{align}
  W_c(\mathbf{n}_c; \mathbf{l}_d) =&\, {\bigg\langle} 
	G_c(c,d) {\bigg\rangle}_{\mathbf{n}_{\bar{d}}} \nonumber \\
	=&\, U_c+  \mathbf{n}_c\cdot \mathbf{V}_c,\label{Gcd0}
\end{align}
with
\begin{equation}
		U_c=1 , \qquad  \mathbf{V}_c= \alpha_c \mathbf{l}_d. \label{UVc0}
\end{equation}
For an initial state polarization $\mathbf{P}_c$ we put $\mathbf{n}_c=\mathbf{P}_c$, 
and get an angular distribution known from the weak hyperon decay 
$\Lambda\rightarrow p\pi^-$ \cite{Okun,GF2}.

The matrix element describing the decay of a polarized $\bar{c}$ (anti)baryon into a polarized 
$\bar{d}$ (anti)baryon is similar to that of Eq.(\ref{HypMel}), 
\begin{equation}
	 {\cal{M}}(\bar{c}\rightarrow \bar{d} \pi)  =
	 \bar{v}(p_{\bar{c}},s_{\bar{c}})(A'+B'\gamma_5)v(p_{\bar{d}},s_{\bar{d}}). \label{HypMelan}
\end{equation}
The relation between the parameters $A,B$ and $A',B'$ is clarified in Refs.\cite{Don1,Don2}.

The square of the anti-baryon matrix element of Eq.(\ref{HypMelan})
is factorized exactly as the baryon-matrix element of Eq.(\ref{MkvadWeak}),
\begin{equation}
	\left| {\cal{M}}({\bar{c}}\rightarrow \bar{d}\pi)\right|^2 = 
	 R({\bar{c}}\rightarrow {\bar{d}}\pi)\, G(\mathbf{n}_{\bar{c}},\mathbf{n}_{\bar{d}}),
\end{equation}
where $\mathbf{n}_{\bar{c}}$ and $\mathbf{n}_{\bar{d}}$ are the spin vectors of
 baryons ${\bar{c}}$ and ${\bar{d}}$ in
their rest systems.

The functions $R({\bar{c}}\rightarrow {\bar{d}}\pi)$ and $G(\mathbf{n}_{\bar{c}},\mathbf{n}_{\bar{d}})$
 are tied to hyperons ${\bar{c}}$ and ${\bar{d}}$ in exactly the same way as 
those tied to hyperons $c$ and $d$, Eqs.(\ref{Rrate}) and (\ref{Gccd}), or to be specific,
\begin{equation}
	G({\bar{c}}, {\bar{d}}) = 1+\alpha_{\bar{c}} \mathbf{n}_{\bar{c}}\cdot \mathbf{l}_{\bar{d}}
	  +\alpha_{\bar{c}} \mathbf{n}_{\bar{d}}\cdot \mathbf{l}_{\bar{d}}
		+\mathbf{n}_{\bar{c}}\cdot \mathbf{L}_{\bar{c}}(\mathbf{n}_{\bar{d}},\mathbf{l}_{\bar{d}} ).
		\label{antiGccd}
\end{equation}

For  CP conserving interactions the asymmetry parameters  of the hyperon pair $c,d$ 
are related to those of anti-hyperon pair $\bar{c},\bar{d}$   by
\begin{equation}
	\alpha_c=-\alpha_{\bar{c}},\quad \beta_c=-\beta_{\bar{c}}, \quad 
	   \gamma_c=\gamma_{\bar{c}}.
\end{equation}

%
%


%
%
\section{Electromagnetic hyperon decays: real photons}\label{EM3}

Electromagnetic transitions such as $\Sigma^0\rightarrow\Lambda\gamma$ and 
$\Xi^0\rightarrow\Sigma^0\gamma$ are readily investigated in $e^+e^-$ annihilation. 
The   electromagnetic $\Sigma^0 \rightarrow \Lambda$ transition  is caused by 
the four-current 
 \cite{Pil} 
\begin{align}
	J_\mu(c\rightarrow d) =&\frac{1}{m_c+m_d} \bigg[ F_1  \bigg\{ \frac{k^2}{m_d-m_c} 
	 \gamma_{\mu}+k_{\mu}\bigg\} \nonumber\\
	&  +F_2i\sigma_{\mu\nu}k^{\nu}\bigg],
\end{align}
with $k=p_c-p_d$. This transition current is gauge invariant, 
meaning $k\cdot J=0$. The $F_1(k^2)$ and $F_2(k^2)$ contributions are each,
 by themselves, gauge invariant.
We shall ignore the $F_1$ term, which vanishes for real photons, $k^2=0$,
 and stay with the $F_2$ term. We denote by $\mu_{cd}$,
\begin{equation}
	\mu_{cd}=eF_2/(m_c+m_d),
\end{equation}
the strength of the magnetic-moment transition.
As a consequence, the expression for the matrix element for any electromagnetic 
$\Sigma^0\rightarrow\Lambda\gamma$ like decay,  becomes 
\begin{align}
	{\cal{M}}_\gamma(c\rightarrow d\gamma)=&
	 \mu_{cd} \bar{u}_d(p_d,s_d) \left( \sigma^{\mu\nu}e_\mu^\star(-ik_\nu)\right) 
	    u_c(p_c,s_c)\nonumber\\
	=&\mu_{cd} \bar{u}_d (p_d,s_d) ( \,/\!\!\!e^\star /\!\!\!k ) u_c(p_c,s_c),
	\label{SLmatrix}
\end{align}
where $s_c$ and $s_d$ are the spin four-vectors of the two baryons.

It is convenient to write the square of this matrix element on the form
\begin{align}
	\left| {\cal{M}}_\gamma(c\rightarrow d\gamma)\right|^2 &=\mu_{cd}^2
	{\rm Tr}\Big[  \half(1+\gamma_5/\!\!\! s_d)(/\!\!\!p_d+m_d)\, / \!\!\!e^\star  \,/  \!\!\!k
\nonumber\\
 & \times (/ \!\!\! p_c+m_c) \half(1+\gamma_5/\!\!\! s_c) \,/\!\!\!e^\star /\!\!\!k 
\Big] \nonumber \\&=H_{\gamma}^{\mu\nu}(k) e_\mu(k)e^{\star}_\nu(k) ,\label{Mkvadrat}
\end{align} 
with $H_{\gamma}^{\mu\nu}(k)$ referred to as the hadron tensor. We have also made use of the
simplifying identity
\begin{equation}
	e^\mu i\sigma_{\mu\nu}k^\nu= - \, / \!\!\!e  \,/  \!\!\!k ,
\end{equation}
valid for real photons.

Summation over the two photon-spin directions entails replacing $e_\mu(k)e^{\star}_{\nu}(k)$ 
by $-g_{\mu\nu}$. This leads to
\begin{equation}
	\sum_{e_\gamma}\left| {\cal{M}}_\gamma(c\rightarrow d\gamma)\right|^2 = 
	 R(c\rightarrow d\gamma)\, G_\gamma(\mathbf{n}_c,\mathbf{n}_d),\label{Mkvadrgam}
\end{equation}
and again  $\mathbf{n}_c$ and $\mathbf{n}_d$ are the spin vectors of baryons $c$ and $d$ in
their rest systems. Photon polarizations are summed over. There are  also electromagnetic 
transitions between charged baryons, but in this section we limit ourselves to electromagnetic 
transitions between
neutral baryons.

The factorization of Eq.(\ref{Mkvadrgam}) is chosen so  that the fractional decay rate
 $R(c\rightarrow d\gamma)$
is the unpolarized part of Eq.(\ref{Mkvadrgam}) and its 
$G_\gamma(\mathbf{n}_c,\mathbf{n}_d)$ factor
the normalized spin-density-distribution function. Here, 
unpolarized means averaged over the spin
directions of both initial and final baryons.
 
The fractional decay rate, $R(c\rightarrow d\gamma)$ of Eq.(\ref{DefBgam}), has the same structure as the 
corresponding one for weak baryon decays,  Eq.(\ref{Rrate}),
\begin{eqnarray}
  R(c\rightarrow d\gamma) &=& 2m_c  \Gamma(c\rightarrow d\gamma)/\Phi(c\rightarrow d\gamma) ,
	 \nonumber\\
	&=& \mu_{cd}^2(m_c^2-m_d^2)^2 ,\label{Rrealph}
 \label{DefBgam}
\end{eqnarray}
where $\Phi(c\rightarrow d\gamma)=\Phi(m_c;m_d,m_\gamma)$ is the phase-space volume.

 The  electromagnetic decay width is
\begin{equation} 
  \Gamma(c\rightarrow d\gamma) = \frac{1}{2\pi} \mu_{cd}^2 \omega^3 ,
	\label{GamDecay}
\end{equation}
where $\omega$ is the photon energy. Remember, that this width 
is obtained after  averaging over both initial and final baryon spin states.

The spin-density-distribution function of Eq.(\ref{Mkvadrgam}) involves an implicit summation 
over photon polarizations. For such a case 
\begin{equation}
	G_\gamma(\mathbf{n}_{c},\mathbf{n}_{d})= 1-
	\mathbf{n}_{c}\cdot \mathbf{l}_{\gamma}\,  \mathbf{l}_{\gamma}\cdot \mathbf{n}_{d},
	\label{Photav}
\end{equation}	
where $\mathbf{l}_{\gamma}$ is a unit vector in the direction of motion of the photon, 
and $ \mathbf{l}_{d}=-\mathbf{l}_{\gamma}$ a unit vector in the direction of motion of
baryon $d$, both in the rest system of baryon $c$.

We notice that when both hadron spins are parallel or anti-parallel to the photon
momentum, then the decay probability vanishes, a property of angular-momentum conservation.
We also notice that expression (\ref{Photav}) cannot be written in the 
$\alpha\beta\gamma$ representation of Eqs.(\ref{Gccd}) 
and (\ref{Gccd_2}). 

When the spin of the final-state baryon $d$ is  not measured, the relevant spin-density  is 
obtained by forming the
 average over the spin directions $\mathbf{n}_d$,
\begin{align}
  W_\gamma(\mathbf{n}_c; \mathbf{l}_d) =&\, {\bigg\langle} 
	G_\gamma(c,d) {\bigg\rangle}_{\mathbf{n}_d} \nonumber \\
	=&\, U_c+  \mathbf{n}_c\cdot \mathbf{V}_c,\label{Gcd01}
\end{align}
with
\begin{equation}
		U_c=1 , \qquad  \mathbf{V}_c= 0. \label{UVc01}
\end{equation}
Thus, the decay-distribution function is independent of the initial-state baryon spin vector $\mathbf{n}_c$.

The anti-particle matrix element corresponding to the particle matrix element
 of Eq.(\ref{SLmatrix}), is simply
\begin{equation}
	{\cal{M}}_\gamma(\bar{c}\rightarrow \bar{d}\gamma)=\mu_{cd}
	\bar{v}_{\bar{c}}(p_{\bar{c}},s_{\bar{c}})
	  ( \,/\!\!\!e^\star /\!\!\!k )
	   v_{\bar{d}} (p_{\bar{d}},s_{\bar{d}}).
	\label{SLmatrix-at}
\end{equation}
We assume the parameter $\mu$ is the same for particle transitions $c\rightarrow d$ 
as for anti-particle transitions $\bar{c}\rightarrow \bar{d}$.

The normalized spin density corresponding to the antiparticle matrix element of 
Eq.(\ref{SLmatrix-at}) is the same as that corresponding to the particle 
matrix element of Eq.(\ref{SLmatrix}), as given in Eq.(\ref{Photav}), 
provided we replace the particle spin vectors $\mathbf{n}_{c}$ and $\mathbf{n}_{d}$ 
by the anti-particle spin vectors $\mathbf{n}_{\bar{c}}$ and $\mathbf{n}_{\bar{d}}$ .

The possibility to search for P-violating admixtures in the electromagnetic 
 decay $\Sigma^0\rightarrow\Lambda\gamma$ was suggested in  Ref.\cite{St1}.
  Such  contributions are created by making the substitution
	\begin{equation}
 / \!\!\!e^*  \,/  \!\!\!k \rightarrow (1-b \gamma_5) / \!\!\!e^*  \,/  \!\!\!k ,
\label{P-sub}
\end{equation}
in the decay amplitude. Moreover, 
if one  can measure  hyperon and anti-hyperon sequential decays
 simultaneously  tests for CP violation become possible.

The substitution (\ref{P-sub}) 
changes the normalized spin density (\ref{Photav}) into 
 \begin{equation}
	G_\gamma(\mathbf{n}_{c},\mathbf{n}_{d})= 1-
	\mathbf{n}_{c}\cdot \mathbf{l}_{\gamma}\,  \mathbf{l}_{\gamma}\cdot \mathbf{n}_{d}
	+ \rho_{c} \left[\mathbf{n}_{c}\cdot \mathbf{l}_{\gamma} -
	\mathbf{n}_{d}\cdot \mathbf{l}_{\gamma}\right],
	\label{Photav-p}
\end{equation}	
with asymmetry parameter
\begin{equation}
 \rho_{c} =\frac{2\Re  (b)}{1+|b|^2}.
\end{equation}
Similarly, the decay width of Eq.(\ref{GamDecay}) is changed into
\begin{equation} 
  \Gamma(c\rightarrow d\gamma) = \frac{1}{2\pi}\,(1+|b|^2) \mu_{cd}^2 \omega^3.
	\label{GamDecay-P}
\end{equation}

Parity violating admixtures in the anti-particle decay
 $\bar{\Sigma}^0\rightarrow\bar{\Lambda}\gamma$ can also be simulated by the substitution 
of Eq.(\ref{P-sub}). For generality we replace $b$ by $\bar{b}$, and simultaneously remark 
that hemiticity requires  $\bar{b}=-b^\star$.
The spin density for the anti-particle decay becomes
\begin{equation}
	G_\gamma(\mathbf{n}_{\bar{c}},\mathbf{n}_{\bar{d}})= 1-
	\mathbf{n}_{\bar{c}}\cdot \mathbf{l}_{\gamma}\,  \mathbf{l}_{\gamma}\cdot \mathbf{n}_{\bar{d}}
	- \rho_{\bar{c}} \left[\mathbf{n}_{\bar{c}}\cdot \mathbf{l}_{\gamma} -
	\mathbf{n}_{\bar{d}}\cdot \mathbf{l}_{\gamma}\right],
	\label{Photav-p-anti}
\end{equation}	
and
\begin{equation}
 \rho_{\bar{c}} =\frac{2\Re(\bar{b})}{1+|\bar{b}|^2}.
\end{equation}
The P-violating interference term now enters with the opposite sign. 
If CP is conserved then  $\bar{b}=-b$. For a full discussion of P and CP 
conservation in this context we refer to Ref.\cite{St1}.
%
%

%
%
\section{Electromagnetic hyperon decays: virtual photons}\label{EM3ee}

The leptonic decay
	$\Sigma^0 \rightarrow \Lambda e^+e^-$
is a small fraction of the electromagnetic decay $\Sigma^0 \rightarrow \Lambda \gamma$ 
\cite{cour,Alff}. The lepton pair of the leptonic decay is interpreted as 
the decay product of a virtual,  
massive photon. This pair is often called    a Dalitz lepton pair. 

The steps to follow in order to find the cross-section distribution for virtual
photons are well known. The square of the reduced matrix element is written as 
  \begin{equation}
	\left| {\cal M}_e(c\rightarrow d e^+e^-) \right|^2 =
	   \frac{1}{m_\gamma^4}H_e^{\mu\nu} L_{\mu\nu} ,
		\label{spin-subst}
\end{equation}
where $H_e^{\mu\nu}$ is the hadron tensor and  $L_{\mu\nu}$ the lepton tensor.

The hadron tensor can be extracted from Eq.(\ref{Mkvadrat}), 
\begin{align}
	H_e^{\mu\nu}(c\rightarrow d e^+e^-)&=\mu_{cd}^2
	{\rm Tr}\Big[  \half(1+\gamma_5/\!\!\! s_d)(/\!\!\!p_d+m_d)\, \sigma^{\mu\tau}k_{\tau}
	\nonumber\\& \times
  (/ \!\!\! p_c+m_c) \half(1+\gamma_5/\!\!\! s_c) \,\sigma^{\nu\lambda}k_{\lambda}\Big].
\label{OMkvadrat}
\end{align}

We need the square of ${\cal M}_e$ averaged over baryon spins but summed  over lepton spins. The summation over 
lepton spins leads to a lepton tensor,
\begin{align}
	L_{\mu\nu}(k_1,k_2) &= e^2\sum_{l\, spin} 
	\bar{v}(k_2)\gamma_\mu u(k_1)\bar{u}(k_1)\gamma_\nu v(k_2) \nonumber\\
	  &= 4e^2\big[ k_\mu k_\nu -k_{1\mu}k_{1\nu} -k_{2\mu}k_{2\nu} - \half g_{\mu\nu}k^2\big].\quad \label{leptonts}
\end{align}

Next, we integrate over the
lepton momenta. For this purpose we rewrite the phase-space element as
\begin{equation}
	   \textrm{dLips}(p_c; p_d, k_1, k_2)=\frac{1}{2\pi}\,\textrm{d}m_\gamma^2\,
		\textrm{dLips}(p_c; p_d, k)\,\textrm{dLips}(k; k_1,k_2)
\end{equation}
with $k^2=m_\gamma^2$ and $\textrm{dLips}(k; k_1,k_2)$ the phase-space element for 
the lepton pair, as in Appendix C. However, care should be excercised since in 
many experiments the efficiency is very sensitive to the lepton momentum.

The integration over the lepton phase space affects only the lepton tensor. Thus, we note that
\begin{equation}
	\left\langle k_{1\mu} k_{1\nu} \right\rangle =\bigg[ \frac{1}{3}(1-\frac{m_e^2}{k^2}) k_\mu k_\nu
	 -\frac{1}{12}\, k^2(1-\frac{4m_e^2}{k^2})g_{\mu\nu}\bigg] \left\langle 1 \right\rangle ,\label{kkave}
\end{equation}
and similarly for $\left\langle k_{2\mu} k_{2\nu} \right\rangle$, with brackets denoting integration over
lepton phase space, $\textrm{dLips}(k; k_1,k_2)$, and $\left\langle 1 \right\rangle $
 denoting the phase-space volume itself. 
The term proportional to $k_\mu k_\nu$ in Eq.(\ref{kkave})  vanishes due to gauge invariance.
As a consequence, we get as  average of the lepton tensor,
\begin{align}
	\left\langle L_{\mu \nu} \right\rangle &= L(k^2)(-g_{\mu \nu} ) \label{Ltot} \\
	L(k^2)&=\alpha_e k^2   \sqrt{1-\frac{4m_e^2}{k^2} }\left[ 1- 
	\frac{1}{3}\left(1-\frac{4m_e^2}{k^2}\right)\right] 
	  . \label{Laverage}
\end{align}

The lepton tensor $L_{\mu\nu}$ of Eq.(\ref{Ltot}) comes with a factor  $(-g_{\mu \nu} )$. 
Contracting it with the hadron tensor 
$H_{\mu\nu}(c\rightarrow dg)$, with $g$ representing  the virtual photon, 
 is equivalent to summing over photon polarizations. We write 
\begin{eqnarray}
\big| {\cal{M}}_e(c\rightarrow d g)\big|^2 & =&- H_{\mu}^{\mu}(c\rightarrow d  g), \nonumber \\
	 &=& R(c\rightarrow d g)\, G(\mathbf{n}_c ,\mathbf{n}_d ). \label{Mvirtgam}
\end{eqnarray}
The factorization is chosen so that $R(c\rightarrow d g)$ is spin independent, 
and so that the spin-independent term of $G(\mathbf{n}_c ,\mathbf{n}_d )$ is unity.

The functions $R$ and $G$ are easily calculated. Neglecting terms unimportant for 
the $\Sigma^0\rightarrow \Lambda\gamma$ transition, we get for the 
fractional decay rate of Eq.(\ref{Mvirtgam}), 
\begin{align}
	R(c\rightarrow d g) &= 2m_c  \Gamma(c\rightarrow dg)/\Phi(c\rightarrow dg) ,
	 \nonumber \\
	&=\mu_{cd}^2 \bigg[  (m_c-m_d)^2-m_\gamma^2\bigg] (m_c+m_d)^2.
\end{align}
where $\Phi(c\rightarrow dg)=\Phi(m_c;m_d,m_\gamma)$ is the phase-space volume.
For $m_\gamma =0$ we recover $R(c\rightarrow d\gamma)$ for real photons, Eq.(\ref{Rrealph}).

Again neglecting terms unimportant for the $\Sigma^0\rightarrow \Lambda\gamma$ transition,
the properly normalized  spin density reads, 
\begin{equation} 
G(\mathbf{n}_c ,\mathbf{n}_d )=
  1 - \mathbf{n}_c \cdot \mathbf{l}_\gamma   \mathbf{l}_\gamma \cdot\mathbf{n}_d .						
\end{equation}
Thus, it is in this  approximation  also equal to  the normalized spin density for
 real photons, Eq.(\ref{Photav}). The exact expressions for $R$ and $G$ 
are given in Appendix E.

Next, we combine the matrix elements for the transitions $c\rightarrow d g$
and $g\rightarrow  e^+e^-$, $g$ representing a virtual photon of mass 
$m_\gamma$.

Since the lepton tensor of Eq.(\ref{Laverage}) lacks spin dependence, 
$G(g\rightarrow  e^+e^- )=1$,  we have the spin-density relation
\begin{equation}
	G(c\rightarrow d e^+e^- )=G(c\rightarrow d g)G(g\rightarrow  e^+e^- ),
\end{equation}
and a corresponding $R$-factor relation
\begin{equation}
	R(c\rightarrow d e^+e^-  )= R(c\rightarrow d g )
	 R(g\rightarrow  e^+e^-  ).
\end{equation}

The function $R(g\rightarrow  e^+e^-)$ collects the remains, 
the lepton tensor of Eq.(\ref{Laverage}) 
multiplied by the propagator $1/k^4$ of Eq.(\ref{spin-subst}),
\begin{equation}
	R(g\rightarrow  e^+e^-  )=\frac{\alpha_e}{k^2} \sqrt{1-\frac{4m_e^2}{k^2} }\left[ 1- 
	\frac{1}{3}\left(1-\frac{4m_e^2}{k^2}\right)\right] . \label{MasDist}
\end{equation}
This expression comes with the phase-space element
\begin{equation}
	\textrm{dLips}=\frac{1}{2\pi}\,\textrm{d}m_\gamma^2\,\, \,
		\textrm{dLips}(p_c; p_d, k),
\end{equation}
where $k$ is the four-momentum of the virtual photon and $k^2=m_\gamma^2$. 
Remember that $m_\gamma^2\geq 4m_e^2$ so there is no singularity in $R(g\rightarrow  e^+e^- )$.

Deviations from the Dalitz-distribution function of Eq.(\ref{MasDist}) signals the importance
form factors in the virtual photon exchange.
%
%
%

\section{Folding}

Our general aim is to calculate the cross-section distributions for $e^+e^-$ annihilation into
  $\Sigma^0\bar{\Sigma}^0$ pairs that subsequently decay, as 
$\Sigma^0\rightarrow\Lambda\rightarrow p$ or
 $\bar{\Sigma}^0\rightarrow\bar{\Lambda}\rightarrow \bar{p}$, 
and as illustrated in Fig.\ 2. The first step in this endeavour is to perform the folding of a 
product of spin densities, a technique  especially adapted to spin one-half baryons.  

\begin{figure}[h]
\begin{center}
      \centerline{\scalebox{0.70}{ \includegraphics{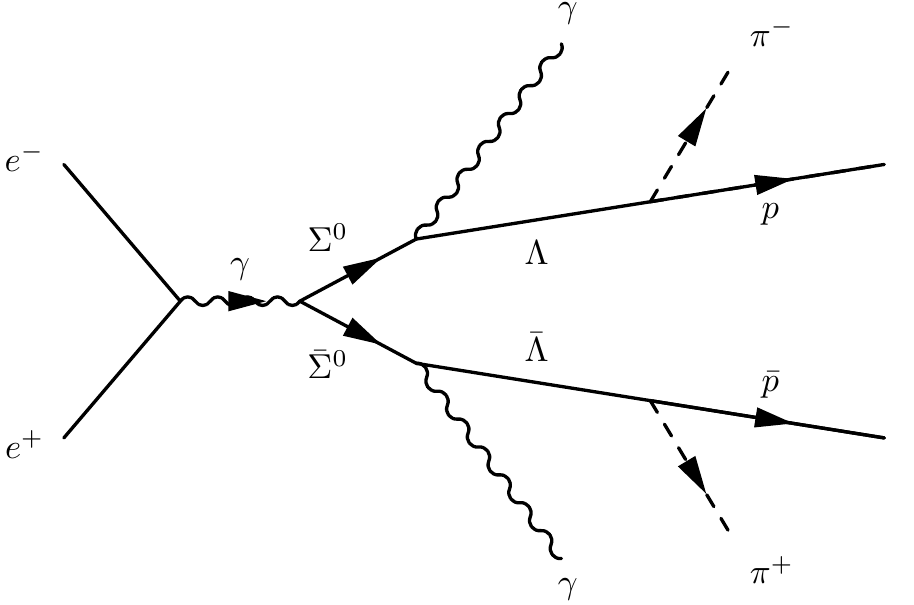} }} 
      \caption{Graph describing the reaction 
$e^+ e^- \rightarrow  \bar{\Sigma}^0 \Sigma^0$, and the subsequent decays, 
$\Sigma^0\rightarrow\Lambda\gamma;\Lambda\rightarrow p\pi^-$  
and $\bar{\Sigma}^0\rightarrow\bar{\Lambda}\gamma;\bar{\Lambda} \rightarrow \bar{p}\pi^+$.
 The reaction graphed can, in addition to photons, 
be mediated by   vector charmonia, such as  $J/\psi$, 
$\psi'$ and $\psi(2{\textrm S})$. Solid lines refer to baryons, dashed to mesons, and wavy to photons.}
\end{center}
\label{F200-fig}
\end{figure}
A folding procedure  implies forming an average over
 intermediate spin directions $\mathbf{n}$ according to the prescription
\begin{equation}
\big{\langle } 1\big{\rangle }_{\mathbf{n}}   =1, \quad 
\big{\langle }  \mathbf{n} \big{\rangle }_{\mathbf{n}}   =0, \quad
\big{\langle }  \mathbf{n}\cdot \mathbf{k}  \mathbf{n}\cdot \mathbf{l}  \big{\rangle }_{\mathbf{n}}   
=\mathbf{k}\cdot \mathbf{l} .\label{Defaverage}
 \end{equation}
For more details see Ref.\cite{GF1}.

In the present case there are five spin densities; the annihilation spin density 
$S(\mathbf{n}_{\Sigma},\mathbf{n}_{\bar{\Sigma}})$ of Eq.(\ref{ISD}); the spin densities of the 
 electromagnetic and weak decays, Eqs.(\ref{Photav}) and (\ref{Gccd}), 
\begin{align}
	G(\Sigma^0\rightarrow \Lambda\gamma) = 1- &
	\mathbf{n}_{\Sigma}\cdot \mathbf{l}_{\gamma}\,  \mathbf{l}_{\gamma}\cdot \mathbf{n}_{\Lambda},
	\label{GSigL}\\
		G(\Lambda \rightarrow p\pi^- ) = 1+& \alpha_{\Lambda} \mathbf{n}_{\Lambda}\cdot \mathbf{l}_{p}
	  +\alpha_{\Lambda} \mathbf{n}_{p}\cdot \mathbf{l}_{p}
	\nonumber \\	& \qquad +\mathbf{n}_{\Lambda}\cdot \mathbf{L}_{\Lambda}(\mathbf{n}_{p},\mathbf{l}_{p} ),
\end{align}	
with $\mathbf{L}_{\Lambda}(\mathbf{n}_{p}, \mathbf{l}_{p})$ defined in Eq.(\ref{Gccd_2});
and the anti-hyperon versions of the last two spin densities. Remember that 
the symbol $\mathbf{l}$ represents a unit vector.

The spin density for the   $\Sigma^0\rightarrow p $ transition is obtained by folding
a product  of spin densities. Averaging over the Lambda and final-state
proton spins, according to the folding prescription Eq.(\ref{Defaverage}), gives us
\begin{align}
G(\Sigma^0\rightarrow p)  &=  {\bigg \langle} G(\Sigma^0\rightarrow \Lambda\gamma ) \, 
        G(\Lambda \rightarrow p \pi^-) {\bigg \rangle}_{\mathbf{n}_\Lambda,\mathbf{n}_p } \nonumber\\
       &= 1-\alpha_{\Lambda} \mathbf{n}_{\Sigma}\cdot \mathbf{l}_{\gamma} 
        \mathbf{l}_{\gamma}\cdot \mathbf{l}_{p}. \label{hyperon_chain}
\end{align}
We notice that this spin density does not depend on the asymmetry parameters 
$\beta_\Lambda$ and $\gamma_\Lambda$, a consequence of the average over the 
final-state-proton-spin directions.

To the baryon decay chain $\Sigma^0\rightarrow\Lambda\rightarrow p$ there is a corresponding anti-baryon
decay chain
 $\bar{\Sigma}^0\rightarrow\bar{\Lambda}\rightarrow \bar{p}$, and a
 corresponding transition-spin density. 

To go  from the baryon to the anti-baryon case, we simply replace the baryon variables by 
their anti-baryon  counterparts, 
$\mathbf{n}_{\Sigma} \rightarrow\mathbf{n}_{\bar{\Sigma}}$,
 $\alpha_{\Lambda} \rightarrow\alpha_{\bar{\Lambda}}$, etc.

The inclusion of parity violation in the $\Sigma^0\rightarrow\Lambda\gamma$ decay is 
straightforward. 
To this end we replace $G(\Sigma^0\rightarrow \Lambda\gamma)$ of Eq.(\ref{GSigL}) by 
\begin{equation}
	G(\Sigma^0\rightarrow \Lambda\gamma) =1- 
	\mathbf{n}_{\Sigma}\cdot \mathbf{l}_{\gamma}\,  \mathbf{l}_{\gamma}\cdot \mathbf{n}_{\Lambda}
	+ \rho_{\Sigma} \left[\mathbf{n}_{\Sigma}\cdot \mathbf{l}_{\gamma} -
	\mathbf{n}_{\Lambda}\cdot \mathbf{l}_{\gamma}\right],
\end{equation}
of Eq.(\ref{Photav-p}), and get	
\begin{align}
G(\Sigma^0\rightarrow p)  &=  {\bigg \langle} G(\Sigma^0\rightarrow \Lambda\gamma ) \, 
        G(\Lambda \rightarrow p \pi^-) 
				{\bigg \rangle}_{\mathbf{n}_\Lambda,\mathbf{n}_p } \nonumber\\
       &=\Big( 1 -\rho_\Sigma \alpha_\Lambda \mathbf{l}_{\gamma}\cdot \mathbf{l}_{p}\Big)
			-\alpha_{\Lambda} \mathbf{n}_{\Sigma}\cdot \mathbf{l}_{\gamma} 
        \mathbf{l}_{\gamma}\cdot \mathbf{l}_{p}. \label{hyperon_chain2}
\end{align}
Therefore, as a consequence of parity violation the normalization of the 
cross-section distribution acquires a small angular dependent term. 

%

%

\section{Single chain decays} 

Single-chain decays of $\Sigma^0$ hyperons can be studied in the $e^+e^-$ annihilation
into $\Sigma^0\bar{\Sigma}^0$ pairs,
provided the $\bar{\Sigma}^0$ is somehow identified, \textit{e.g.}, as a missing hyperon
 Ref.\cite{GF4}.
 The spin-density state 
of the $\Sigma^0$  will then be
obtained from Eq.(\ref{ISD}) as,
\begin{equation}
	{\bigg\langle} S(\mathbf{n}_{\Sigma},\mathbf{n}_{\bar{\Sigma}}) 
  {\bigg\rangle}_{\mathbf{n}_{\bar{\Sigma}}}=
	{\cal{R}}  
	    + {\cal{S}}\, \mathbf{N}\cdot \mathbf{n}_{\Sigma} . \label{One_ch}
	\end{equation}
	
A $\Sigma^0$ hyperon in a state of polarization $\mathbf{P}_\Sigma$, subject to the condition 
$|\mathbf{P}_\Sigma |\leq 1$,  is characterized by a normalized
spin-density function, 
\begin{equation}
	S_\Sigma(\mathbf{n}_\Sigma)= 1+\mathbf{P}_\Sigma\cdot \mathbf{n}_\Sigma. \label{SP}
\end{equation}
Therefore, by Eq.(\ref{One_ch}) the hyperon polarization is in  the present  case equal to 
 $\mathbf{P}_\Sigma={\cal{S}}\, \mathbf{N}/{\cal{R}}$.

If a $\Sigma^0$ hyperon of polarization $\mathbf{P}_\Sigma$ 
undergoes an electromagnetic decay, $\Sigma^0\rightarrow \Lambda \gamma $,
we can determine the spin-density distribution of the $\Lambda$ hyperon by 
folding the initial state $\Sigma^0$  spin density of Eq.(\ref{SP}) with the $\Sigma^0$ 
decay distribution of Eq.(\ref{Photav}), to get
\begin{align}
	 W_\Lambda(\mathbf{n}_\Lambda;\mathbf{l}_\Lambda )= &\,{\bigg\langle} S_\Sigma(\mathbf{n}_\Sigma)
	G_\gamma(\mathbf{n}_\Sigma,\mathbf{n}_\Lambda) {\bigg\rangle}_{\mathbf{n}_\Sigma}
	\nonumber\\
	= &\,1-  \mathbf{P}_\Sigma\cdot \mathbf{l}_\Lambda\, \mathbf{l}_\Lambda\cdot \mathbf{n}_\Lambda,\label{Wlam}
\end{align}
with $\mathbf{l}_\Lambda=-\mathbf{l}_\gamma$ and a $\Lambda$ polarization 
\begin{equation}
	\mathbf{P}_\Lambda= -\mathbf{P}_\Sigma\cdot \mathbf{l}_\Lambda\, \mathbf{l}_\Lambda.
\end{equation}
Consequently,   the $\Lambda$ polarization is directed along the $\Lambda$  
momentum $\mathbf{l}_\Lambda$,
 a fact which is independent of the initial $\Sigma^0$ hyperon spin.

Let us now consider also the weak decay of the $\Lambda$-hyperon, $\Lambda\rightarrow p \pi^-$,
which is described by the spin density $ G_\Lambda(\mathbf{n}_\Lambda, \mathbf{n}_p)$ of 
Eq.(\ref{Gccd}).
Since the spin of the final-state proton is usually not measured, 
we form the average over the proton spin directions. The spin-density-distribution function 
of Eq.(\ref{Wlam}) is now expanded to 
\begin{align}
	 W_p(\mathbf{l}_\Lambda,  \mathbf{l}_p )= &\,{\bigg\langle} S_\Sigma(\mathbf{n}_\Sigma)
	G_\gamma(\mathbf{n}_\Sigma,\mathbf{n}_\Lambda)
	G_p(\mathbf{n}_\Lambda,\mathbf{n}_p)
{\bigg\rangle}_{\mathbf{n}_\Sigma,\mathbf{n}_\Lambda,\mathbf{n}_p }
	\nonumber\\
	= &\,1- \alpha_\Lambda \mathbf{P}_\Sigma\cdot \mathbf{l}_\Lambda\, 
	\mathbf{l}_\Lambda\cdot \mathbf{l}_p,
	\nonumber\\
	= &\,1+ \alpha_\Lambda \mathbf{P}_\Lambda\cdot \mathbf{l}_p.
	\label{W2lam}
\end{align}

The decay chain $\Sigma^0\rightarrow \Lambda\gamma \rightarrow p\pi^-$ makes part of our
annihilation process and it is therefore of interest to investigate what
 additional information may 
be obtained by measuring the spin of the final-state proton. Thus, instead of the 
spin density of Eq.(\ref{hyperon_chain2}) we investigate the spin density
\begin{align}
	G(\Sigma^0\rightarrow p )= {\bigg \langle}	G(\Sigma^0\rightarrow \Lambda \gamma)
  \,		G(\Lambda \rightarrow p\pi^- )
	{\bigg \rangle}_{{\mathbf{n}}_\Lambda} .
\end{align}
 Invoking the vector-function identity of Eq.(\ref{Gccd_2}) we get  
\begin{align}
	G(\Sigma^0\rightarrow p) = 1+ & \alpha_{\Lambda} \mathbf{n}_{p}\cdot \mathbf{l}_{p} \nonumber \\
	-&\mathbf{n}_{\Sigma}\cdot \mathbf{l}_{\gamma}  \bigg[ 
	  \alpha_{\Lambda} \mathbf{l}_{\gamma}\cdot \mathbf{l}_p 
	          + \mathbf{n}_{p}\cdot {\mathbf{L}}_{\Lambda}(\mathbf{l}_{\gamma}, -\mathbf{l}_{p})
	  \bigg].\label{DecSigma}
\end{align}

Finally, the spin-density-distribution function for  the final state proton is obtained as
\begin{align}
	S(\mathbf{n}_{p})=& {\bigg \langle} S(\mathbf{n}_{\Sigma})	S(\mathbf{n}_{\Sigma}, \mathbf{n}_{p} )\,	
	{\bigg \rangle}_{\mathbf{n}_\Lambda,\mathbf{n}_\Sigma }\nonumber\\
	 =&{U}_p + \mathbf{V}_p\cdot \mathbf{n}_{p} ,\\
	 U_p=& 1 - \alpha_\Lambda \mathbf{P}_\Sigma\cdot \mathbf{l}_\gamma 
	       \mathbf{l}_\gamma \cdot\mathbf{l}_p , \\
	   \mathbf{V}_p=&   \alpha_\Lambda  \, \mathbf{l}_p- 
	       \mathbf{P}_\Sigma\cdot \mathbf{l}_\gamma 
	         {\mathbf{L}}_\Lambda( \mathbf{l}_\gamma ,-\mathbf{l}_p ) .
\end{align}
This result describes a proton polarization which is $\mathbf{V}_p/U_p$. It is 
explicitly dependent on $\alpha_\Lambda $, but there is a hidden dependence on 
$\beta_\Lambda $ and $\gamma_\Lambda $ in the vector function $\mathbf{L}_\Lambda$.

%
 %
%
%
\section{Production and  decay of $\Sigma^0 \bar{\Sigma}^0 $ pairs}
%
%

Now, we come to the main task of our investigation; 
production and  decay of $\Sigma^0 \bar{\Sigma}^0 $ pairs. The starting point 
is the reaction $e^+e^-\rightarrow \Sigma^0 \bar{\Sigma}^0$, 
the spin-density distribution  of which was calculated in Sect.3. 
We name it $S(\mathbf{n}_{\Sigma},\mathbf{n}_{\bar{\Sigma}})$. 
The explicit expression is given by Eq.(\ref{ISD}), with $\mathbf{n}_1, \mathbf{n}_2$ 
replaced by $\mathbf{n}_{\Sigma},\mathbf{n}_{\bar{\Sigma}}$. 

The  spin-density distribution
 $W_{\Sigma}(\mathbf{n}_{{\Sigma}},\mathbf{n}_p)$ for the  decay chain 
$\Sigma^0\rightarrow \Lambda \gamma; \Lambda\rightarrow p\pi^-$ is given in
 Eq.(\ref{DecSigma}).
We write
\begin{align}
	W_\Sigma(\mathbf{n}_\Sigma,\mathbf{n}_p) &= U_\Sigma+  \mathbf{n}_\Sigma\cdot 
	\mathbf{V}_\Sigma, \label{Yachain}\\
	U_\Sigma &=       1+  \alpha_{\Lambda} \mathbf{n}_{p}\cdot \mathbf{l}_{p}             \\
	\mathbf{V}_\Sigma &=- \mathbf{l}_{\gamma}  \left[ 
	  \alpha_{\Lambda} \mathbf{l}_{\gamma}\cdot \mathbf{l}_p 
	          + \mathbf{n}_{p}\cdot {\mathbf{L}}_{\Lambda}(\mathbf{l}_{\gamma}, -\mathbf{l}_{p})
	  \right],
			\label{Yabarchain}
\end{align}
and ditto for $W_{\bar{\Sigma}}(\mathbf{n}_{\bar{\Sigma}},\mathbf{n}_{\bar{p}})$.
We are only interested in decay chains of $\Sigma^0$ and $\bar{\Sigma}^0$
which are each others anti chains.

The final-state-angular distributions are obtained by folding the
spin distributions for production and decay, according to presciption (\ref{Defaverage}). 
Invoking Eq.(\ref{ISD}) for the production step and 
 Eqs.(\ref{Yachain}) and its anti-distribution for the  decay steps, we get the 
angular distribution
\begin{align}
	 W_{\Sigma\bar{\Sigma}}(\mathbf{l}_a)=& \,{\bigg\langle}
	     S(\mathbf{n}_\Sigma, \mathbf{n}_{\bar{\Sigma}})   
	 	W_\Sigma(\mathbf{n}_\Sigma,\mathbf{n}_p)W_{\bar{\Sigma}}
		(\mathbf{n}_{\bar{\Sigma}},\mathbf{n}_{\bar{p}})
	 	{\bigg\rangle}_{\mathbf{n}_\Sigma, \mathbf{n}_{\bar{\Sigma}}} \nonumber \\
	 =& \,  {\cal{R}} U_\Sigma U_{\bar{\Sigma} }
	    + {\cal{S}}U_{\bar{\Sigma} }\, \mathbf{N}\cdot \mathbf{V}_{\Sigma}
	+ {\cal{S}}U_\Sigma\,\mathbf{N}\cdot \mathbf{V}_{\bar{\Sigma}} \nonumber \\
	& + {\cal{T}}_1 
	  \mathbf{V}_{\Sigma} \cdot \hat{\mathbf{p}}\mathbf{V}_{\bar{\Sigma}} \cdot \hat{\mathbf{p}}
	    + {\cal{T}}_2  \mathbf{V}_{\Sigma\bot} \cdot \mathbf{V}_{\bar{\Sigma}\bot} \nonumber \\ &
		+ {\cal{T}}_3 \mathbf{V}_{\Sigma\bot} \cdot \hat{\mathbf{k}}\mathbf{V}_{\bar{\Sigma}\bot}
		\cdot \hat{\mathbf{k}}
		\nonumber\\
			&  + {\cal{T}}_4  \bigg( \mathbf{V}_{\Sigma} \cdot \hat{\mathbf{p}}
			\mathbf{V}_{\bar{\Sigma}\bot} \cdot \hat{\mathbf{k}} 
			 + \mathbf{V}_{\bar{\Sigma}} \cdot \hat{\mathbf{p}}\mathbf{V}_{\Sigma\bot} \cdot \hat{\mathbf{k}} 
			  \bigg)  ,\label{Hdef2}
\end{align}
where $\mathbf{l}_a$ denotes the ensemble of $\mathbf{l}$ values in the decays.

The angular distributions of Eq.(\ref{Hdef2}) still depend on the spin vectors
 $\mathbf{n}_p$ and $\mathbf{n}_{\bar{p}}$ which are difficult to measure. If we 
are willing to consider 
the spin averages, then variables $U$ and $\mathbf{V}$ simplify,
\begin{align}
		U_{\Sigma}&=1 , \qquad  \mathbf{V}_{\Sigma}= -\alpha_{\Lambda}\, \mathbf{l}_{\Lambda}\cdot 
		\mathbf{l}_p\mathbf{l}_{\Lambda} \nonumber \\
		U_{\bar{\Sigma}}&=1 , \qquad  \mathbf{V}_{\bar{\Sigma}}=
		-\alpha_{\bar{\Lambda}}\, \mathbf{l}_{\bar{\Lambda}}\cdot 
		 \mathbf{l}_{\bar{p}}\mathbf{l}_{\bar{\Lambda} } . \label{UVSig}
\end{align}
Since $U_{\Sigma}=U_{\bar{\Sigma}}=1$ the effect of the folding is to make, in 
the spin-density function $S(\mathbf{n}_{\Sigma},\mathbf{n}_{\bar{\Sigma}})$ of Eq.(\ref{ISD}), 
 the replacements 
$\mathbf{n}_\Sigma\rightarrow  \mathbf{V}_\Sigma$ and 
$\mathbf{n}_{\bar{\Sigma}}\rightarrow  \mathbf{V}_{\bar{\Sigma}}$.
We notice that the $U$ and $\mathbf{V}$ variables are independent of the
weak-asymmetry parameters $\beta_\Lambda$ and $\gamma_\Lambda$. 
Their dependence is hidden in the vector function 
${\mathbf{L}}_{\Lambda}(\mathbf{l}_{\gamma}, -\mathbf{l}_{p})$ of 
Eq.(\ref{Yabarchain}), and which is absent in Eq.(\ref{Hdef2}).

Inserting the expressions of Eq.(\ref{UVSig}) into the spin-density function 
of Eq.(\ref{Hdef2}) we get
\begin{align}
	 W_{\Sigma\bar{\Sigma}}(\mathbf{l}_a)
	 =& \,  {\cal{R}} 
	    -\alpha_{\Lambda} {\cal{S}}\, \mathbf{N}\cdot \mathbf{l}_\Lambda\,  \mathbf{l}_\Lambda \cdot \mathbf{l}_p
	-\alpha_{\bar{\Lambda}} {\cal{S}}\,\mathbf{N}\cdot \mathbf{l}_{\bar{\Lambda}}\,  
	           \mathbf{l}_{\bar{\Lambda}} \cdot \mathbf{l}_{\bar{p}}
	 \nonumber \\
	& + \alpha_{\Lambda}\alpha_{\bar{\Lambda}}\,   \mathbf{l}_{\Lambda}\cdot \mathbf{l}_p
	  \mathbf{l}_{\bar{\Lambda}} \cdot \mathbf{l}_{\bar{p}}
	\, \bigg{[}\,  {\cal{T}}_1                          
	   \mathbf{l}_\Lambda \cdot \hat{\mathbf{p}}
		                 \mathbf{l}_{\bar{\Lambda}} \cdot \hat{\mathbf{p}} \nonumber\\
	   & + {\cal{T}}_2  \mathbf{l}_{{\Lambda}\bot} \cdot \mathbf{l}_{\bar{\Lambda}\bot} 
	 + {\cal{T}}_3 \mathbf{l}_{\Lambda\bot} \cdot \hat{\mathbf{k}}\mathbf{l}_{\bar{\Lambda}\bot}
	\cdot \hat{\mathbf{k}}
		\nonumber\\
			&  + {\cal{T}}_4  \bigg( \mathbf{l}_{\Lambda} \cdot \hat{\mathbf{p}} 
			\mathbf{l}_{\bar{\Lambda}\bot} \cdot \hat{\mathbf{k}} 
			 + \mathbf{l}_{\bar{\Lambda}} \cdot \hat{\mathbf{p}}\mathbf{l}_{\Lambda\bot} \cdot \hat{\mathbf{k}} 
			  \bigg) \bigg{]} .\label{Hdef3a}
\end{align}
Thus, this is the angular distribution obtained when  folding the product of spin densities 
for production and decay. 

%
%
%
\section{Differential distributions}

Explicit expressions for the structure 
functions ${\cal{R}}$, ${\cal{S}}$, and ${\cal{T}}$ 
are given in Appendix B. With their help we can rewrite the differential distribution
 function of  Eq.(\ref{Hdef3a}) as
\begin{equation}
\begin{split}
{\cal{W}}({\boldsymbol{\xi}})=&\ \bigg[ {\cal{F}}_0+{{{\eta}}}{\cal{F}}_1\bigg]
     \\ 
		&-\sqrt{1-\eta^2}\sin(\Delta \Phi)\sin\theta \cos\theta\ 
		\bigg[ \alpha_\Lambda{\cal{F}}_2{\cal{F}}_5  +   
		 \alpha_{\bar{\Lambda}}{\cal{F}}_3{\cal{F}}_6
		\bigg]           \\&             
   +\alpha_\Lambda \alpha_{\bar{\Lambda}}{\cal{F}}_2{\cal{F}}_3\ 
	{\bigg[} (\eta+\cos^2\!\theta)\, 
	{\cal{F}}_4 -{{{\eta}}}\sin^2\!\theta\, {\cal{F}}_7 \\ &
	  \qquad +(1+\eta)\sin^2\!\theta\,{\cal{F}}_8\\ & \qquad+
	  \sqrt{1-{{\eta}}^2}\cos({{\Delta\Phi}})\sin\theta \cos\theta\,  {\cal{F}}_9  \bigg]
	    , \label{Weqn:pdf}
\end{split}
\end{equation}
where the argument ${\boldsymbol{\xi}}$ of the angular functions is a nine-dimensional vector ${\boldsymbol{\xi}}=(\theta,\Omega_{\Lambda},\Omega_{p},\Omega_{\bar{\Lambda}},\Omega_{\bar{p}})$.

The  ten angular 
functions ${\cal{F}}_k({\boldsymbol{\xi}})$  are defined as;
\begin{align}
	{\cal{F}}_0({\boldsymbol{\xi}}) =&1, \nonumber\\
	{\cal{F}}_1({\boldsymbol{\xi}}) =&{\cos^2\!\theta},\nonumber\\
	{\cal{F}}_2({\boldsymbol{\xi}}) =&\mathbf{l}_{\Lambda}\cdot \mathbf{l}_p ,\nonumber\\
  {\cal{F}}_3({\boldsymbol{\xi}}) =&\mathbf{l}_{\bar{\Lambda}}\cdot \mathbf{l}_{\bar{p}},\nonumber\\        
  {\cal{F}}_4({\boldsymbol{\xi}}) =&\mathbf{l}_{\Lambda}\cdot \hat{\mathbf{p}}
                                  \mathbf{l}_{\bar{\Lambda}}\cdot \hat{\mathbf{p}},  \nonumber\\
	{\cal{F}}_5({\boldsymbol{\xi}}) =& \mathbf{N}\cdot \mathbf{l}_\Lambda,  \nonumber\\
	{\cal{F}}_6({\boldsymbol{\xi}}) =& \mathbf{N}\cdot \mathbf{l}_{\bar{\Lambda}} 	,\nonumber\\
	{\cal{F}}_7({\boldsymbol{\xi}}) =&\mathbf{l}_{\Lambda\bot} \cdot \mathbf{l}_{\bar{\Lambda}\bot} 	   ,\nonumber\\
	{\cal{F}}_8({\boldsymbol{\xi}}) =&   \mathbf{l}_{\Lambda\bot} \cdot \hat{\mathbf{k}}
		                             \mathbf{l}_{\bar{\Lambda}\bot} \cdot \hat{\mathbf{k}} 
																/\sin^2\theta
								 ,\nonumber\\
  {\cal{F}}_9({\boldsymbol{\xi}}) =& \left( \mathbf{l}_{\Lambda}\cdot \hat{\mathbf{p}}
			\mathbf{l}_{\bar{\Lambda}\bot} \cdot \hat{\mathbf{k}} +
			\mathbf{l}_{\bar{\Lambda}} \cdot \hat{\mathbf{p}}
			\mathbf{l}_{{\Lambda} \bot}
			\cdot \hat{\mathbf{k}} \right) /\sin\theta. \label{TenF}
\end{align}             
The cross-section distribution (\ref{Hdef3a}), and also the ten angular functions above, 
 depend on a number of unit vectors; 
$\hat{\mathbf{p}}$ and $-\hat{\mathbf{p}}$ are unit vectors along the directions of motion of the $\Sigma^0$ 
and the $\bar{\Sigma}^0$  in the c.m. system;  
$\hat{\mathbf{k}}$  and $-\hat{\mathbf{k}}$ are unit vectors along the directions of motion of the incident electron 
and positron in the c.m. system;
$\mathbf{l}_{\Lambda}$ and $\mathbf{l}_{\bar{\Lambda}}$ are unit vectors along the 
directions of motion of the $\Lambda$ and $\bar{\Lambda}$ in the rest systems of the $\Sigma^0$ 
and the $\bar{\Sigma}^0$; 
$\mathbf{l}_{p}$ and $\mathbf{l}_{\bar{p}}$ are unit vectors along the 
directions of motion of the $p$ and the $\bar{p}$ in the rest systems of the $\Lambda$ 
and the $\bar{\Lambda}$. Longitudinal  and transverse components of vectors are defined with respect 
 to the $\hat{\mathbf{p}}$ direction.

The differential distribution function ${\cal{W}}({\boldsymbol{\xi}})$ of Eq.(\ref{Weqn:pdf}) involve two parameters related to the $e^+e^-\to\Sigma^0\bar{\Sigma}^0$ 
reaction that can be determined by data: the ratio of form factors $\eta$,  and the relative phase of form factors $\Delta\Phi$. 
In addition, the distribution function ${\cal{W}}({\boldsymbol{\xi}})$ depends on the
 weak-asymmetry parameters 
$\alpha_\Lambda$ and $\alpha_{\bar{\Lambda}}$ of the two Lambda-hyperon decays.
The dependence on the weak-asymmetry parameters $\beta$ and $\gamma$ drops out, since
 final-state-proton 
and anti-proton spins are not measured.

An important conclusion to be drawn from the differential distribution of Eq.(\ref{Weqn:pdf}) is 
that when the phase $\Delta \Phi$ is small, the parameters $\alpha_\Lambda$ and $\alpha_{\bar{\Lambda}}$
 are strongly correlated and therefore difficult 
to separate. In order to contribute to the experimental precision of 
$\alpha_\Lambda$ and $\alpha_{\bar{\Lambda}}$ a non-zero value of $\Delta \Phi$ is required.

The sequential differential decay distribution of a single-tagged $\Sigma^0$ produced 
in $e^+e^-$ annihilation can be obtained form Eq.(\ref{Weqn:pdf})
by suitably integrating over the angular vaiables $\Omega_{\bar{\Lambda}}$ and $\Omega_{\bar{p}}$. 
As a result we get the differential distribution for $\Sigma^0$  production and decay,
\begin{align} 
\rd \sigma &\propto \, \bigg[ 
{\cal{R}} 
	    -\alpha_{\Lambda} {\cal{S}}\, \mathbf{N}\cdot \mathbf{l}_\Lambda\,  \mathbf{l}_\Lambda \cdot \mathbf{l}_p \bigg] \, \rd \Omega\, \rd \Omega_\Lambda \rd \Omega_p \nonumber
     \\& 	=    \,
 \bigg[ 1+\eta\cos^2\!\theta
    - \alpha_\Lambda\sqrt{1-\eta^2}\sin(\Delta \Phi)\sin\theta \cos\theta\,\nonumber\\& 
		\qquad \times \cos\theta_{\Lambda p}
		\sin\theta_{\Lambda }	\sin\phi_{\Lambda }
          \bigg]\,\rd \Omega\,  \rd \Omega_\Lambda \rd \Omega_p.  \label{Wprim:pdf}
\end{align}
Here, $\theta$ is the $\Sigma^0$ production angle, $\theta_{\Lambda p}$ the relative angle between 
the vectors $\mathbf{l}_{\Lambda}$ and $\mathbf{l}_{p}$, and $\theta_{\Lambda}$ and 
$\phi_{\Lambda}$ the directional angles of $\mathbf{l}_{\Lambda}$ in the global
coordinate system of \ref{Ang_coord}. From the
angular distribution of Eq.(\ref{Wprim:pdf}) we can determine the product $\alpha_\Lambda \sin(\Delta \Phi)$, 
and from the corresponding $\bar{\Sigma^0}$ distribution the product 
$\alpha_{\bar{\Lambda}} \sin(\Delta \Phi)$.

%
%
\section{Cross-section distributions}

We shall now consider the phase-space imbedding of the differential-distribution
function of Eq.(\ref{Weqn:pdf}). We start with the cross-section-distribution function
for creation of a pair of baryons, $e^+e^-\rightarrow B\bar{B}$. 
Combining Eqs.(\ref{Basicsigma}),
 (\ref{TwoDsig}), and (\ref{Hintro}),
we get
\begin{equation}
	\rd \sigma(e^+e^-\rightarrow B\bar{B}) =\frac{p}{k} \frac{\alpha_e^2 \,D(s)}{4s^2}\,S(\mathbf{n}_{b},\mathbf{n}_{\bar{b}})\, 
	{\rd \Omega},\label{Compton}
\end{equation}
where $\Omega$  are the baryon scattering angles in the c.m. system.

Next we consider the propagator factors associated with the sequential decays of the baryons 
 $\Sigma^0$ and $\bar{\Sigma}^0$ produced in the $e^+e^-$ annihilation process.
These sequential decays are illustrated in Fig.2.
 There are three factors 
associated with the square of each propagator. Let us consider the decay  $c\rightarrow dg$, where 
$g$ can represent a pion or a photon. 
Other decay modes are also possible to incorporate. Then, we have
\begin{equation}
	{\cal{P}}_c=\Big[ \frac{\pi}{m_c\Gamma_c}\delta(s_c-m_c^2)\Big]
	\Big[ \frac{\rd s_c}{2\pi}\, \textrm{dLips}(p_c;p_d,p_g)\Big]
	\Big[ R_c G_c\Big].
\end{equation}
Here, the first factor comes from squaring the propagator in the Feynman diagram; the second factor 
from dividing the phase-space element into a product of two-body phase-space elements; 
and the third factor is the reduced matrix element squared for the decay $c\rightarrow dg$, 
and the product of the normalized spin density $G_c$ and  
the fractional decay rate  $R_c$.

The fractional decay rate  $R_c$ is defined in Eq.(\ref{DefBgam}) as
\begin{equation}
  R(c\rightarrow dg) = 2m_c  \Gamma(c\rightarrow dg)/\Phi(c\rightarrow dg) ,
 \label{DefBgamprim}
\end{equation}
where $\Phi$ is the two-body phase-space volume, and $\Gamma(c\rightarrow dg)$ the channel width
for the decay $c\rightarrow dg$. It was defined to be spin averaged for both 
initial and final baryon  states. However, in a sequential decay both final 
 spin-state contribution must  be included. This is achieved by multiplying 
$R(c\rightarrow d\gamma)$ by a factor of two. This factor can be incorporated in
the channel width $\Gamma(c\rightarrow dg)$, reinterpreting it to include the sum over
final baryon spin states. Finally, we observe that
\begin{equation}
	\textrm{dLips}(p_c;p_d,p_g)=\Phi_c(c\rightarrow dg) \frac{\rd \Omega_c}{4\pi},
\end{equation}
giving as a consequence a  ${\cal{P}}$ factor
\begin{equation}
	{\cal{P}}_c=G_c \,  \frac{\Gamma(c \rightarrow dg)}{\Gamma(c\rightarrow all)}\, 
	\frac{\rd \Omega_c}{4\pi},
\end{equation}
with $\Omega_c$ the angular variable in the rest system of baryon $c$. 
In our application the indices $c$ and $\bar{c}$ are representatives for 
 $c=\Sigma^0,\Lambda$ and $\bar{c}=\bar{\Sigma}^0,\bar{\Lambda}$
 
The differential-distribution function ${\cal{W}}({\boldsymbol{\xi}})$ of Eq.(\ref{Weqn:pdf})
is obtained by {\bfseries folding}  a product of spin densities
\begin{equation}
	{\cal{W}}({\boldsymbol{\xi}})={\bigg \langle} S(\mathbf{n}_b,\mathbf{n}_{\bar{{b}}})
	 \prod_{c,\bar{c}} G_c(\mathbf{n}_c,\mathbf{n}_d )
	{\bigg \rangle}_{\mathbf{n}} .\label{Pplus:decay}
\end{equation}
Folding involves averages over spin directions, but as remarked, cross-section 
distributions requires 
summing over the spin directions. Thus, an average over the spin density 
$S(\mathbf{n}_b,\mathbf{n}_{\bar{{b}}})$ is accompanied by an extra factor of four,
and it is not normalized to unity either but to ${\cal{R}}$.

The folding formula Eq.(\ref{Pplus:decay}) combined with Eqs.(\ref{Compton}) and
 (\ref{Compton-cc})
gives  the {\bfseries master equation}
\begin{equation}
	\rd \sigma =\rd \sigma(e^+e^-\rightarrow\Sigma^0\bar{\Sigma}^0)\, 
	\frac{\cal{W}({\boldsymbol{\xi}})}{\cal{R}}
	\prod_{c,\bar{c}}
	  \Bigg[ \frac{\Gamma(c\rightarrow dg)}{\Gamma(c\rightarrow all)}\frac{\rd \Omega_c}{4\pi}\Bigg].
	 \label{GammaCross2}
\end{equation}
This readily understood structure agrees with that found in Ref.\cite{GF1}.

Since spin densities are normalized, except for the annihilation density
 $ S(\mathbf{n}_b,\mathbf{n}_{\bar{{b}}})$, the overall normalization condition reads 
\begin{equation}
	\int {\cal{W}} ({\boldsymbol{\xi}}) \prod_{c,\bar{c}} \frac{\rd \Omega_c}{4\pi}=   {\cal{R}}.
\end{equation}
This normalization  is checked explicitely in single-chain-sequential decay in Ref.\cite{GF4}.

When the $\Sigma^0\rightarrow\Lambda\gamma$ reaction is involved, and the photon is real, 
 then the
channel width in Eq.(\ref{GammaCross2}) is for all practical purposes equal to the total width, 
$\Gamma(\Sigma^0\rightarrow\Lambda\gamma)=\Gamma(\Sigma^0\rightarrow all)$. 

We also point out that
for virtual photons the cross-section distribution of Eq.(\ref{GammaCross2}) receives  
an additional lepton factor,
\begin{equation}
	\frac{1}{2\pi}\,\textrm{d}m_\gamma^2\,\, \,
		R(g\rightarrow  e^+e^-  ),
\end{equation}
where $m_\gamma$ is the virtual photon mass, and $R$ the Dalitz function
\begin{equation}
	R(g\rightarrow  e^+e^-  )=\frac{\alpha_e}{k^2} \sqrt{1-\frac{4m_e^2}{k^2} }\left[ 1- 
	\frac{1}{3}\left(1-\frac{4m_e^2}{k^2}\right)\right] . \label{MasDistEnd}
\end{equation}

The $e^+e^-$ annihilation reactions described above are all concerned with annihilation 
through ordinary photons, as illustrated in Fig.\ref{F1-fig}. However, the same reactions can  be initiated
 by other vector mesons as well. 
Of special interest is the $J/\psi$ case, which is treated in Ref.\cite{GF3}, and 
which is accessible to the BESIII experiment. By making the replacement 
\begin{equation}
	\frac{\alpha_e^2}{s^2}\rightarrow \frac{\alpha_\psi \alpha_g}{(s-m_\psi^2)^2+m_\psi^2\Gamma(m_\psi)}
\end{equation}
 in the photon-induced reaction, Eq.(\ref{Compton}), we get the cross-section-distribution formula for 
annihilation through the $J/\psi$ meson. The meaning of the parameters $\alpha_\psi$ and $\alpha_g$
is explained in Ref.\cite{GF3}. This is equivalent to replacing in the master formula of 
Eq.(\ref{GammaCross2}), the photon-induced  $e^+e^-\rightarrow\Sigma^0\bar{\Sigma}^0$
annihilation cross section  by the  $J/\psi$  induced cross section.

%
%
\appendix
%
%
\section{Graph calculation}

In this Appendix we shall work out the phase-space density for the two-step case.
Our notation follows Pilkuhn \cite{Pil}. The cross-section distribution can be
 written as 
\begin{equation}
	\rd \sigma= \frac{1}{2\sqrt{\lambda(s,m_e^2,m_e^2)}} \, \overline{|{\cal{M}}|^2}\,
	   \textrm{dLips}(k_1+k_2;\{\mathbf{l}_i\},\{\mathbf{l}_{i'}\})	  ,
\end{equation}	   
where $\{\mathbf{l}_i\}$ are the final-state momenta in the hyperon decay chain and $\{\mathbf{l}_{i'}\}$ 
are the final-state momenta in the anti-hyperon decay chain.
The average over the squared matrix element indicates summation over final-state spins 
and average over initial-state lepton.  The definitions of the
particle momenta are explained in Fig.1.

Since $\Gamma\ll M$ for the intermediate propagators, their squares may be 
approximated as 
\begin{equation}
	\frac{1}{(s-M^2)^2 +M^2\Gamma^2(\sqrt{s})}= \frac{\pi}{M\Gamma(M)}\delta(s-M^2). 
\end{equation}
This makes it convenient to pull out a factor ${\cal{K}}$ from the squared matrix element, 
\begin{equation}
{\cal{K}}=	\prod_i \frac{1}{(s_i-M_i^2)^2 +M_i^2\Gamma_i^2(M_i)},
  \label{Kdef}
\end{equation}
and plug it into the phase-space density. In Eq.(\ref{Kdef}) the product runs over 
the four intermediate-state hyperons.

After some manipulations we can write the modified phase-space density as
\begin{align}
	 {\cal{K}}&\textrm{dLips}(k_1+k_2;\{\mathbf{l}_i\},\{\mathbf{l}_{i'}\})=
	\bigg[ \frac{p}{(4\pi)^2\sqrt{s}} \, \rd \Omega \bigg]_\textrm{CM} \nonumber \\ 
		 & \qquad \quad  \times\prod_i \bigg[ \frac{q_i}{8\pi M_i^2 \Gamma_i(M_i) } \, \rd \Omega_i \bigg]_\textrm{Y},
		 \label{Ph_space_form}
\end{align}
where index CM refers to  the two-body reaction $e^+e^-\rightarrow Y\bar{Y}$, and index Y
to each of the four intermediate-state hyperon decays, in their respective hyperon 
rest systems.
%
%
%
\section{Structure functions}\label{AppC}

The six structure functions ${\cal{R}}$, ${\cal{S}}$, and ${\cal{T}}$ of Eq.(\ref{ISD})  
 depend on the scattering angle $\theta$, in the c.m.\ system, the ratio function $\eta(s)$, and
the phase function $\Delta\Phi(s)$. To be specific \cite{GF2,GF3},
\begin{eqnarray}
{\cal{R}} &=& 1 +\eta  \cos^2\!\theta, \label{DefR}\\
  {\cal{S}} &=& \sqrt{1-\eta^2}\sin\theta\cos\theta\sin(\Delta\Phi), \label{DefS}\\
	{\cal{T}}_1 &=& \eta + \cos^2\!\theta, \\
	{\cal{T}}_2 &=& -\eta\sin^2\!\theta,  \\
	{\cal{T}}_3 &=& 1+\eta, \\
	{\cal{T}}_4 &=& \sqrt{1-\eta^2}\cos\theta\cos(\Delta\Phi). \label{RSTdef}
\end{eqnarray}
The parameters $\eta$ and $\Delta\Phi$ are defined in Eqs.(\ref{alfa_def}) and (\ref{DPHI_def}). 


\section{Phase-space volume}\label{AppD}

The Lorentz invariant two-body phase-space element is by definition
\begin{equation}
	\textrm{dLips}(k; k_1,k_2) = \frac{\rd^3k_1}{(2\pi)^32\omega_1} \frac{\rd^3k_2}{(2\pi)^32\omega_2}
	 (2\pi)^4\delta(k-k_1-k_2).
\end{equation}
Integration exploiting the delta functions leads to
\begin{equation}
	\int \textrm{dLips}(k; k_1,k_2) = \frac{k_c}{4\pi\sqrt{s}}\frac{\textrm{d}\Omega_c}{4\pi}
	 \label{FasVol}
\end{equation} 
where $\sqrt{s}=M$, $k_c$ the momentum, and $\Omega_c$ the  angular variable, 
both in the  c.m.\ system.
In terms of the mass variables
 \begin{equation}
	k_c^2 = \frac{1}{4M^2} \, \bigg[   (M^2+m_1^2-m_2^2)^2-4M^2m_1^2 \, \bigg].
\end{equation} 
 The phase-space volume $\Phi$ is obtained from Eq.(\ref{FasVol}) by integration 
over ${\textrm{d}\Omega_c}$,
\begin{equation}
	\Phi(M;m_1,m_2) = \frac{k_c}{4\pi\sqrt{s}}.
\end{equation}
For equal masses $m_1=m_2=m$ the value of the phase-space volume becomes 
\begin{equation}
	\Phi(M;m,m)\equiv \left\langle 1 \right\rangle               
	=\frac{1}{8\pi} \sqrt{1-\frac{4m^2}{M^2}  }.
\end{equation}

\section{Decay into virtual gamma}\label{AppE}

The squared matrix element $\big| {\cal{M}}(c\rightarrow d g)\big|^2$ for the decay of
 a baryon $c$ into a baryon $d$ and a 
virtual gamma $g$ of mass $m_\gamma$ is given in Eq.(\ref{Mvirtgam}). 
It can be factorized into factors $R(c\rightarrow d g)$ and $G_g (\mathbf{n}_c ,\mathbf{n}_d )$.
The exact expression for the fractional width is,  
\begin{equation}
	R(c\rightarrow d g)=\mu_{cd}^2 \Big[ (m_c-m_d)^2-m_\gamma^2\Big]\Big[(m_c+m_d)^2+\half m_\gamma^2\Big],
\end{equation}
with $2m_e\leq m_\gamma \leq (m_c-m_d)$. In the limit $m_\gamma =0$ we recover $R( c\rightarrow d\gamma)$ for real photons, Eq.(\ref{Rrealph}). The exact expression for the normalized spin density is,
\begin{align}
G_g (\mathbf{n}_c ,\mathbf{n}_d )=&
1 +  B \mathbf{n}_c \cdot \mathbf{l}_\gamma   \mathbf{l}_\gamma \cdot\mathbf{n}_d
       +C\mathbf{n}_c \cdot\mathbf{n}_d , \\
			A=& (m_c + m_d)^2 +\half m_\gamma^2 ,
			            \nonumber\\
			B=&-(m_c+m_d)^2/A , \nonumber \\
			C=&\half m_\gamma^2/A.						
\end{align}
Here, we can without qualm put $B=-1$ and $C=0$.
In this limit we recover the normalized spin density for real photons, Eq.(\ref{Photav}).

\section{Angular functions}\label{Ang_coord}


The cross-section distribution (\ref{Hdef3a}) is a function of two hyperon unit vectors: 
$\mathbf{l}_{\Lambda}$, the direction of motion of the Lambda 
hyperon in the rest system of the Sigma hyperon, and 
$\mathbf{l}_{p}$ the direction of motion of the proton 
 in the rest system of the Lambda hyperon. Plus the corresponding vectors for the 
anti-hyperon chain. In order to handle these
vectors we introduce a common global coordinate system, which we define as follows.
%
%
%
%
%



The scattering plane  of the reaction 
$e^+e^-\rightarrow\Sigma^0\bar{\Sigma}^0$ is spanned by the unit vectors 
 $\hat{\mathbf{p}}=\mathbf{l}_{\Sigma}$ and $\hat{\mathbf{k}}=\mathbf{l}_{e}$, as 
measured in  the c.m.\ system. 
 The scattering plane makes up 
the $xz$-plane, with the $y$-axis  along the normal to the scattering plane. 
We choose a right-handed coordinate system with basis vectors 
	\begin{eqnarray}
	\mathbf{e}_z  &=&  \hat{\mathbf{p}}, \nonumber\\
	\mathbf{e}_y  &=& \frac{1}{\sin\theta } ( \hat{\mathbf{p}}\times \hat{\mathbf{k}} ) ,
	\nonumber \\
	\mathbf{e}_x  &=& \frac{1}{\sin\theta } (\hat{\mathbf{p}}\times \hat{\mathbf{k}} ) 
	 \times\hat{\mathbf{p}}.\label{xunity}
\end{eqnarray}
Expressed in terms of them the initial-state lepton momentum becomes
\begin{equation}
	\hat{\mathbf{k}}= \sin\theta\,  \mathbf{e}_x +\cos\theta\,	\mathbf{e}_z  .
\end{equation}

This coordinate system is used for defining the directional angles of 
the  Lambda and the  proton. The  directional angles of the Lambda hyperon 
in the Sigma hyperon rest system are,
\begin{equation}
	\mathbf{l}_\Lambda=(\cos \phi_\Lambda \sin \theta_\Lambda,  \sin \phi_\Lambda \sin \theta_\Lambda, 
	\cos \theta_\Lambda),
\end{equation}
whereas the  directional angles of the proton in the Lambda hyperon rest system are 
\begin{equation}
	\mathbf{l}_p=(\cos \phi_p \sin \theta_p,  \sin \phi_p \sin \theta_p, \cos \theta_p).
\end{equation}
And so for the anti-hyperons.

An event of the reaction 
$e^+e^-\rightarrow \Sigma^0 \bar{\Sigma}^0 $; $\Sigma^0\rightarrow \Lambda\rightarrow p$;
$\bar{\Sigma}^0\rightarrow \bar{\Lambda}\rightarrow \bar{p}$;
 is specified by a nine-dimensional vector 
${\boldsymbol{\xi}}=(\theta,\Omega_{\Lambda},\Omega_{p},\Omega_{\bar{\Lambda}},\Omega_{\bar{p}}).$ 
The differential-cross-section distribution is proportional to a function 
${\cal{W}}({\boldsymbol{\xi}})$,
which according to Eq.(\ref{Weqn:pdf}) can be decomposed as
\begin{equation}
\begin{split}
{\cal{W}}({\boldsymbol{\xi}})=&\ \bigg[ {\cal{F}}_0({\boldsymbol{\xi}})+{{{\eta}}}{\cal{F}}_1({\boldsymbol{\xi}})\bigg]
     \\ 
		&-\sqrt{1-\eta^2}\sin(\Delta \Phi)\sin\theta \cos\theta\ 
		\bigg[ \alpha_\Lambda{\cal{F}}_2({\boldsymbol{\xi}}){\cal{F}}_5({\boldsymbol{\xi}}) \\&
		 \qquad + \alpha_{\bar{\Lambda}}{\cal{F}}_3({\boldsymbol{\xi}}){\cal{F}}_6({\boldsymbol{\xi}}) 
		\bigg]           \\&             
   +\alpha_\Lambda \alpha_{\bar{\Lambda}}{\cal{F}}_2({\boldsymbol{\xi}}){\cal{F}}_3({\boldsymbol{\xi}})
   \ 
	{\bigg[} (\eta+\cos^2\!\theta)\, 	{\cal{F}}_4({\boldsymbol{\xi}})
	\\& \qquad-{{{\eta}}}\sin^2\theta\, {\cal{F}}_7({\boldsymbol{\xi}}) 
	+(1+\eta)\sin^2\!\theta\,{\cal{F}}_8({\boldsymbol{\xi}})
	\\ &
	   \qquad+
	\  \sqrt{1-{{\eta}}^2}\cos({{\Delta\Phi}})\sin\theta \cos\theta\,  {\cal{F}}_9({\boldsymbol{\xi}})  \bigg]
	    , \label{eqn:pdf}
\end{split}
\end{equation}
The set of ten angular functions, ${\cal{F}}_0({\boldsymbol{\xi}})-{\cal{F}}_9({\boldsymbol{\xi}})$, are defind in Eq.(\ref{TenF}).
 The scalar products  needed for their determination are as follows:
%
%
\begin{align}
	\mathbf{N}\cdot \mathbf{l}_\Lambda
	 =& \sin \theta_\Lambda \sin \phi_\Lambda , \nonumber \\
	 \mathbf{l}_{\Lambda}\cdot \mathbf{l}_p =&
	   \sin\theta_\Lambda\sin\theta_p\cos(\phi_\Lambda-\phi_p)+
              \cos\theta_\Lambda\cos\theta_p \nonumber \\
	   \mathbf{l}_\Lambda\cdot \hat{\mathbf{p}}  =&\cos\theta_\Lambda,
	\nonumber\\
		\mathbf{l}_{\Lambda\bot} \cdot \hat{\mathbf{k}} =& \sin\theta \sin \theta_\Lambda \cos\phi_\Lambda 
		,\nonumber \\
		\mathbf{l}_{\Lambda\bot} \cdot \hat{\mathbf{p}} =& 0, \nonumber \\
		\mathbf{l}_{\Lambda\bot} \cdot \mathbf{l}_{\bar{\Lambda}\bot} =& \sin\theta_\Lambda \sin \theta_{\bar{\Lambda}} 
		\cos(\phi_\Lambda -\phi_{\bar{\Lambda}}).
		  \label{Hdef3} 
\end{align}
We understand that the remaining scalar products are obtained from those above by the substitution 
$(\Lambda;p)\rightarrow (\bar{\Lambda};\bar{p})$. 
With the scalar products of Eq.(\ref{Hdef3} ) in hand
one quickly determines the ten angular functions 
 ${\cal{F}}_k({\boldsymbol{\xi}})$ of  Eq.(\ref{TenF}),
\begin{align}
	{\cal{F}}_0({\boldsymbol{\xi}}) =&1, \nonumber\\
	{\cal{F}}_1({\boldsymbol{\xi}}) =&{\cos^2\!\theta},\nonumber\\
	{\cal{F}}_2({\boldsymbol{\xi}}) =&\sin\theta_\Lambda \sin\theta_p\cos(\phi_\Lambda-\phi_p)+
              \cos\theta_\Lambda\cos\theta_p ,\nonumber\\
{\cal{F}}_3({\boldsymbol{\xi}}) =&\sin\theta_{\bar{\Lambda}}\sin\theta_{\bar{p}}
	  \cos(\phi_{\bar{\Lambda}}-\phi_{\bar{p}})+
            \cos\theta_{\bar{\Lambda}}\cos\theta_{\bar{p}} ,\nonumber\\             
{\cal{F}}_4({\boldsymbol{\xi}}) =&\cos\theta_{\Lambda} \cos\theta_{\bar{\Lambda}},  \nonumber\\
	{\cal{F}}_5({\boldsymbol{\xi}}) =& {\sin\theta_\Lambda}\ \sin\phi_\Lambda,  \nonumber\\
	{\cal{F}}_6({\boldsymbol{\xi}}) =& \sin\theta_{\bar{\Lambda}} \sin\phi_{\bar{\Lambda}} 	,\nonumber\\
	{\cal{F}}_7({\boldsymbol{\xi}}) =&\sin\theta_\Lambda \sin\theta_{\bar{\Lambda}} 
	               \cos(\phi_\Lambda -\phi_{\bar{\Lambda}}) 	,\nonumber\\
		{\cal{F}}_8({\boldsymbol{\xi}}) =&  \sin\theta_{\Lambda}\cos\phi_{\Lambda}
		            \sin\theta_{\bar{\Lambda}} \cos\phi_{\bar{\Lambda}} ,\nonumber\\
			{\cal{F}}_9({\boldsymbol{\xi}}) =& \cos\theta_\Lambda \sin\theta_{\bar{\Lambda}} \cos \phi_{\bar{\Lambda}}  
			+\sin\theta_{\Lambda}\cos \phi_{\Lambda} \cos\theta_{\bar{\Lambda}} .
\end{align}

The differential distribution of Eq.~(\ref{eqn:pdf}) involves two parameters related to the $e^+e^-\to\Sigma^0
\bar{\Sigma}^0$ reaction that can be determined by data: the ratio of form factors $\eta$,  and the relative phase of form factors $\Delta\Phi$. 
In addition, the distribution function ${\cal{W}}({\boldsymbol{\xi}})$ depends on the weak-decay parameters 
$\alpha_\Lambda$ and $\alpha_{\bar{\Lambda}}$ of the two $\Lambda$ hyperon decays.
The dependence on the weak decay parameters $\beta$ and $\gamma$ drops out, since final-state proton 
and anti-proton spins are not measured.

%

%
%
\section*{Acknowledgments}

We have greatly benefitted from discussions with Patric Adlarsson, Annele Heikkil\"a, 
 Andrzej Kupsc, and Stefan Leupold.

%
%

%
%
%

\begin{thebibliography}{99}

\bibitem{Ablikim17a}  M.~Ablikim { et~al.}~(BESIII), Phys.\ Rev.\ D  {\bf 95},  052003 (2017).
\bibitem{pacetti}S.~Pacetti, R.~Baldini~Ferroli and E.~Tomasi-Gutafsson, Phys. Rep. \textbf{550-551} 1 (2015).
\bibitem{Nature} M.~Ablikim et al.\ (BESIII), \textit{Polarization and Entanglement in 
  Baryon-Antibaryon Pair Production in Electron-Positron Annihilation}, arXiv[hep-ex]:1808.08917 (2018).
\bibitem{GF2} G.~ F\"aldt, Eur.\ Phys.\ J.\ A  {\bf 52},  141 (2016).
\bibitem{GF3} G.~ F\"aldt and A.Kupsc, Phys.\ Lett.\ B {\bf 772},  16 (2017).
\bibitem{GF4} G.\ F\"aldt, Phys.\ Rev.\ D   {\bf 97},  053002 (2018).
\bibitem{punjabi}V.~Punjabi, C.~F.~Perdrisat and M.~K.~Jones, Eur. Phys. J. A \textbf{51}, 79 (2015).
\bibitem{St1} S.~S.\ Nair, E.~Perotti, and S.~Leupold, Phys.\ Lett.\ B {\bf 788},  535 (2019).
\bibitem{Okun} L.B.\ Okun, \textit{Leptons and Quarks}, North-Holland, Amsterdam,  1982.
\bibitem{Ber} R.E.\ Behrends,  Phys.\ Rev. {\bf 111},  1691 (1958). 
\bibitem{GF1} G.~ F\"aldt, Eur.\ Phys.\ J.\ A  {\bf 51},  74 (2015).
\bibitem{Pil} H.\ Pilkuhn, \textit{Relativistic Particle Physics} (Springer-Verlag,
  Berlin, 1979).
\bibitem{Dub} A.~Z.~ Dubni\u{c}kova, S.~Dubni\u{c}ka, and M.~P.~ Rekalo, Nuoco Cimento A 
  {\bf 109}, 241 (1996).
 \bibitem{Lee} T.\ D.\ Lee and C.\ N.\ Yang,  Phys.\ Rev.\    {\bf 108},  1645 (1957). 
\bibitem{cour} H.\ Courant et al., Phys.\ Rev.\ Lett.  {\bf 10}, 409 (1963).
\bibitem{Alff} C.\ Alff et al., Phys.\ Rev.  {\bf 137}, B 1105 (1965).
\bibitem{Don1} John F.\ Donoghue and Sandip Pakvasa, Phys.\ Rev.\ Lett.\ {\bf 55},  162 (1985).
\bibitem{Don2} John F.\ Donoghue, Xiao-Gang and Sandip Pakvasa, Phys.\ Rev.\ D{\bf 34},  833 (1986).
\end{thebibliography}
\end{document}